\documentclass[APA,LATO1COL]{WileyNJD-v2}

\pdfoutput=1

\articletype{Research Article}

\usepackage{bm}
\usepackage{textgreek}
\usepackage{xcolor}
\usepackage{derivative}
\usepackage{amsfonts}
\usepackage[FIGTOPCAP]{subfigure}
\usepackage{bm}
\newcommand{\tran}{{\mkern-1.5mu\mathsf{T}}}

\definecolor{RED}{named}{red}

\begin{document}

\title{Reinforcement learning for robust dynamic metabolic control}

\author[1]{Sebastián Espinel-Ríos}
\author[2]{River Walser}
\author[3]{Dongda Zhang}

\authormark{ESPINEL-RÍOS \textsc{et al.}}
\titlemark{Reinforcement learning for robust dynamic metabolic control}

\address[1]{\orgdiv{Biomedical Manufacturing Program}, \orgname{Commonwealth Scientific and Industrial Research Organisation}, \orgaddress{\state{Victoria}, \country{Australia}}}

\address[2]{\orgname{Basis Independent Brooklyn}, \orgaddress{\state{New York}, \country{United States}}}

\address[3]{\orgdiv{Department of Chemical Engineering}, \orgname{University of Manchester}, \orgaddress{\state{Manchester}, \country{United Kingdom}}}

\corres{Corresponding author: Sebastián Espinel-Ríos, \email{sebastian.espinelrios@csiro.au}}

\abstract[Abstract]{
\normalfont{{{Dynamic metabolic control allows key metabolic fluxes to be modulated in real time, enhancing bioprocess flexibility and expanding available optimization degrees of freedom. This is achieved, e.g., via targeted modulation of metabolic enzyme expression.}} However, identifying optimal dynamic control policies is challenging due to {the generally high-dimensional solution space and the need to manage metabolic burden and cytotoxic effects arising from inducible enzyme expression}. {The task is further complicated by stochastic dynamics, which reduce bioprocess reproducibility. We propose a reinforcement learning framework} to derive optimal policies by allowing an agent {(the controller)} to interact with a surrogate dynamic model. {To promote robustness,} we apply domain randomization, enabling the controller to generalize across uncertainties. {When transferred to an experimental system, the agent can in principle continue fine-tuning the policy}. Our {framework} provides an alternative to conventional model-based control such as model predictive control, {which requires model differentiation with respect to decision variables; often impractical for} complex stochastic, nonlinear, stiff, {and} piecewise-defined dynamics. In contrast, our approach relies on forward integration of the model, {thereby simplifying the task}. {We demonstrate the framework in two \textit{Escherichia coli} bioprocesses: dynamic control of acetyl-CoA carboxylase for fatty-acid synthesis and of adenosine triphosphatase for lactate synthesis.}}}


\maketitle

\section{Introduction}
\label{sec:introduction_motivation}
Advanced bioprocessing often involves engineering cellular metabolic networks to introduce new pathways or optimize the efficiency of existing ones \citep{volk_metabolic_2023}. This is typically achieved through genetic engineering strategies, such as inserting, deleting, downregulating, or overexpressing metabolic enzymes. Because metabolic enzymes catalyze reactions within cells, their concentrations directly determine reaction rates. Thus, modulating enzyme expression offers a targeted means for precise control over metabolic fluxes \citep{gao_bifunctional_2024, ruiz_optogenetic_2025, pouzet_promise_2020, komera_bifunctional_2022, wang_development_2022}, thereby making it possible to maximize production efficiency in bioprocesses by reconfiguration of metabolic networks.

{There are two distinct paradigms regarding modulation of enzyme expression: static and dynamic control \citep{brockman_dynamic_2015,hartline_dynamic_2021}.} Static control involves a constant induction level, offering operational simplicity but {hindering} dynamic optimization. Such fixed expression levels are often engineered by selecting suitable promoters, adjusting gene copy numbers{, or by maintaining a constant induction signal. Static control policies can, in principle, be simpler to derive, even through trial and error, yet at the expense of process flexibility and adaptability.} In contrast, under ideal conditions, dynamic metabolic control continuously and reversibly modulates enzyme expression {aided by, e.g., genetic circuits}, enabling greater operational flexibility and providing access to a broader range of metabolic modes {throughout the process}. 

Higher enzyme expression levels do not necessarily translate into increased production efficiency. {That is,} \textit{excessive} enzyme expression can rapidly arrest growth by depleting cellular resources or causing unintended cytotoxic effects (cf. e.g., \citep{ohkubo_hybrid_2024,hoffman_balancing_2025}). {Therefore,} a central challenge in {dynamic} metabolic control is identifying optimal enzyme modulation {trajectories} that maximize product pathway efficiency while minimizing cytotoxicity and intrinsic metabolic burdens. {Another challenging task in dynamic metabolic control is, given an apriori identified optimal intracellular dynamic trajectory (e.g., following a golden batch), how to optimally steer the cell to follow that enzyme expression profile efficiently and consistently.}

 {Elucidating optimal dynamic policies is, however, significantly challenging due to biosystems' and bioprocesses' inherent} nonlinearities (e.g., steep activation and deactivation kinetics), multi-scale dynamics, delayed responses, piecewise or switch-like functions triggered by specific cellular events, and the presence of system uncertainties (e.g., stochastic gene expression, process variability, and external disturbances) \citep{zhang_multiscale_2006, oyarzun_design_2015, glass_nonlinear_2021, olsson_robustness_2022, pal_living_2024}. Consequently, dynamic metabolic control represents a nontrivial \textit{nonlinear}, \textit{stochastic}, and \textit{dynamic} control problem. To address this challenge, we propose reinforcement learning (RL) \citep{sutton_reinforcement_2018, dong_introduction_2020}, a machine-learning-based feedback control approach, to derive optimal dynamic policies for enzyme expression regulation, aiming to maximize production efficiency {and process compliance} under uncertainty.

Traditional model-based dynamic control strategies, such as model predictive control (MPC), rely on derivative-based optimization techniques that require explicit mathematical models. For example, necessary optimality conditions such as the Karush–Kuhn–Tucker (KKT) conditions involve differentiating the model with respect to the decision variables \citep{rawlings_model_2020}. However, mathematical models for bioprocesses and metabolic systems often exhibit highly nonlinear, stiff, or piecewise dynamics, posing significant differentiation challenges and potentially hindering solver convergence in model-based control approaches. Moreover, conventional MPC typically assumes deterministic system dynamics. Although stochastic MPC formulations have been proposed \citep{heirung_stochastic_2018}, their practical implementation remains computationally demanding, particularly for practitioners without specialized expertise in control theory.

In contrast, our proposed RL-based approach enables the generation of robust dynamic metabolic control policies without requiring differentiation of the model with respect to decision variables, as in MPC. Instead, RL learns optimal policies by directly interacting with a {surrogate} environment (or process), which involves integrating the dynamic model forward in time; a task that can be handled in a generally efficient way using {standard} off-the-shelf numerical solvers. In our method, the RL agent (or controller) determines the dynamic metabolic control policies for regulating enzyme expression by maximizing the expected value of a user-defined {objective (or return metric)} that quantifies the biosystem’s production efficiency. {This objective can be tailored to either an economic or a reference tracking control task}.

Additionally, our approach explicitly incorporates system uncertainties into deterministic models through domain randomization, thereby better capturing realistic bioprocess conditions and behavior. This is achieved by exposing the RL controller to varying levels of uncertainty during training, allowing it to learn policies that are not only optimal but also robust to intra- and extracellular disturbances. Moreover, domain randomization provides a systematic framework for evaluating the robustness of different dynamic metabolic control architectures \textit{in silico}, offering a cost-effective and safe environment for early-stage decision-making in bioprocess and dynamic metabolic engineering development.

{In summary, while different strategies have been proposed in the literature to implement dynamic metabolic control to maximize bioproduction, e.g., relying on model-based optimal and predictive control \citep{gadkar_estimating_2005, chang_nonlinear_2016, jabarivelisdeh_adaptive_2020, espinelrios_toward_2024, espinel-rios_experimentally_2024, espinel-rios_hybrid_2024}, here we aim to offer an alternative strategy using RL. Our framework is especially useful when model-based control is difficult to implement (e.g., due to highly nonlinear models, steep piecewise functions, etc.) and when incorporating uncertainty awareness is important to enhance robustness.}

{We demonstrate our approach using two representative case studies. The first one involves the} dynamic metabolic control of acetyl-CoA carboxylase (ACC), a key enzyme regulating fatty acid biosynthesis in \textit{Escherichia coli} \citep{ohkubo_hybrid_2024}. Since intracellular ACC accumulation can lead to cytotoxic effects, precise dynamic modulation is essential to maintain high production efficiency and cell viability. {The second one deals with the dynamic metabolic control of adenosine triphosphatase (ATPase) in a lactate fermentation by \textit{E. coli} \citep{espinel-rios_experimentally_2024}. In this bioprocess, modulating the expression of ATPase unlocks a tunable trade-off between increased specific lactate formation and decreased specific biomass growth. We focus on efficiently tracking a user-defined ATPase dynamic profile, which can be relevant, e.g., when dealing with the task of maximizing the compliance with respect to \textit{golden batches}.} We evaluate {the performance} and robustness of {the} RL-derived dynamic control policies by benchmarking them against static metabolic control policies under varying levels of system uncertainty.

{The remainder of this paper is structured as follows. Section \ref{sec:RL_domain_rand} outlines the proposed RL framework for robust dynamic metabolic control. Section \ref{sec:biological_system} introduces the metabolic systems used as case studies. These systems are then used in Section \ref{sec:control_results} to demonstrate the capabilities of our framework.}

\section{Robust dynamic metabolic control}
\label{sec:RL_domain_rand}
\begin{figure} [htb!]
    \centering
    \includegraphics[scale=0.45]{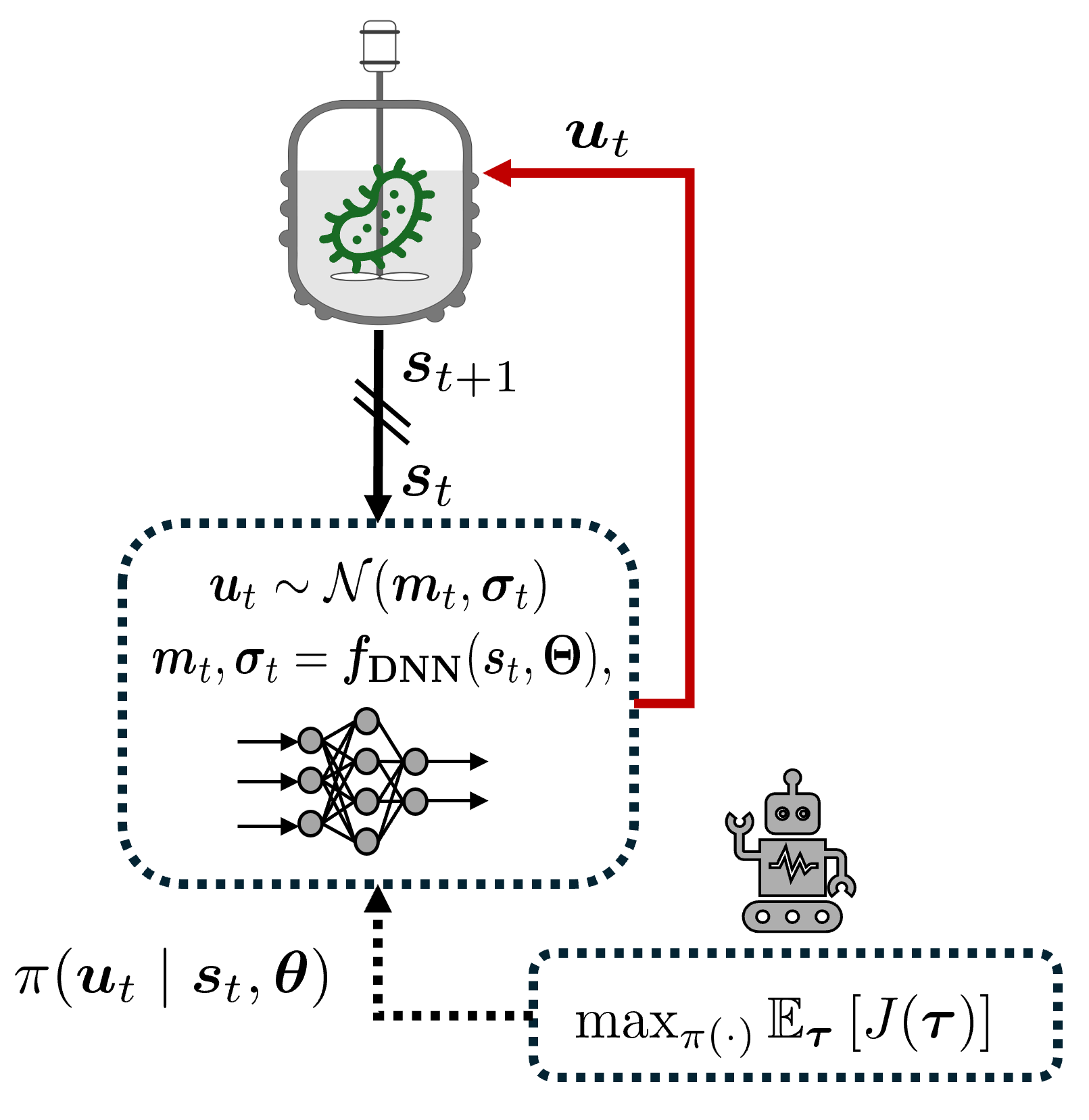}
    \caption{Overview of {the RL framework for robust dynamic metabolic control}. {$\bm{u}_t$: action/input at time $t$; $\bm{s}_t$: featurized system representation at time $t$; $\pi(\cdot)$: stochastic policy; $\bm{m}_t, \bm{\sigma}_t$: mean and standard deviation of Gaussian policy at time $t$; $\bm{\tau}$: joint trajectory of observed states, actions, and rewards; $J(\bm{\tau})$: return over $\bm{\tau}$; $\bm{\theta}$: policy parameters; $\bm{\Theta}$: deep neural network (DNN) parameters}.
    }
    \label{fig:overview}
\end{figure}

{The overall RL framework, which will be the subject of this section, is illustrated in Fig. \ref{fig:overview}}. For generality, we define the system state at discrete time $t$ as the state vector $\bm{x}_t \in \mathbb{R}^{n_x}$. The system evolves to the next discrete time step $\bm{x}_{t+1}$ according to a Markov decision process:
\begin{equation}
    \bm{x}_{t+1} = \bm{f_x}(\bm{x}_t, \bm{u}_t, \bm{\omega}, \bm{d}_t), \quad \forall t \in \{0, 1, \dots, N_x-1\}, \label{eq:dynamics_stoc}
\end{equation}
where $\bm{f_x}: \mathbb{R}^{n_x} \times \mathbb{R}^{n_u} \times \mathbb{R}^{n_\omega} \times \mathbb{R}^{n_d} \to \mathbb{R}^{n_x}$ represents the state transition function. The variables $\bm{u}_t \in \mathbb{R}^{n_u}$, $\bm{\omega} \in \mathbb{R}^{n_{\omega}}$, and $\bm{d}_t \in \mathbb{R}^{n_d}$ denote the metabolic control input, constant model parameters, and stochastic disturbances of both intracellular and extracellular origin, respectively. When the disturbances follow a probability distribution $\mathcal{P}_d(\bm{d}_t)$, Eq. \eqref{eq:dynamics_stoc} describes the stochastic system dynamics. The state transition occurs over $N_x$ time intervals, with $\bm{x}_0$ representing the initial condition.

{For the sake of generality,} we define a trajectory $\bm{\tau}$ as the collection of states, inputs, and rewards over the entire process:
\begin{equation} 
    \bm{\tau} = \{  (\bm{x}_0, \bm{u}_0, R_1, \bm{x}_1), (\bm{x}_1, \bm{u}_1, R_2, \bm{x}_2), \ldots,  (\bm{x}_{N_x-1}, \bm{u}_{N_x-1}, R_{N_x}, \bm{x}_{N_x}) \},
\end{equation}
where $R_t \in \mathbb{R}$ represents the system reward at time $t$, a user-defined metric that quantifies system performance in response to the applied control input.

The control input $\bm{u}_t$ is sampled from a policy distribution $\pi(\cdot)$, which is conditioned on a feature vector representation of the system $\bm{s}_t \in \mathbb{R}^{n_s}$, providing contextual information about the system state at time $t$. That is:
\begin{equation}
    \bm{u}_{t} \sim \pi(\bm{u}_t \mid \bm{s}_t, \bm{\theta}),
\end{equation}
where $\bm{\theta} \in \mathbb{R}^{n_\theta}$ represents the policy parameters, which define the shape and properties of the policy distribution.

The conditional probability of a trajectory $\bm{\tau}$, given the policy parameters $\bm{\theta}$, is expressed as:
\begin{equation}
\mathrm{P}(\bm{\tau} \mid \bm{\theta}) = \mathrm{P}(\bm{x}_0)  \prod_{t=0}^{N_x-1} \left[ \pi(\bm{u}_t \mid \bm{s}_t, \bm{\theta})  \mathrm{P}(\bm{x}_{t+1} \mid \bm{x}_t, \bm{u}_t) \right],
\label{eq:prob_dis_tau}
\end{equation}
where $\mathrm{P}(\bm{x}_0)$ represents the probability of the initial state and $\mathrm{P}(\bm{x}_{t+1} \mid \bm{x}_t, \bm{u}_t)$ represents the state transition probability given the current state and applied control input.

\subsection{Policy gradients}
\label{subsec:pol_grad}
The primary objective in RL is to maximize the expected system performance, represented by the return function $J(\bm{\tau})$, where the expectation is taken over trajectories $\bm{\tau}$ generated under the policy $\pi(\cdot)$:
\begin{equation} 
\max_{\pi(\cdot)} \mathbb{E}_{\bm{\tau} \sim \mathrm{P}(\bm{\tau} \mid \pi)} \left[ J(\bm{\tau}) \right]. \label{eq:rl_problem} 
\end{equation}
In the context of dynamic metabolic control, $J(\cdot)$ can represent key performance metrics in bioprocessing, such as the final product titer or volumetric productivity, which are common economic objectives. {$J(\cdot)$ can also be tailored to represent a reference tracking problem, effectively \textit{minimizing} the tracking error. Examples of these will be outlined along with the case studies.}

To solve the stochastic dynamic optimization problem in \eqref{eq:rl_problem}, we apply gradient ascent to iteratively update the policy parameters:
\begin{equation}
\bm{\theta}_{m+1} = \bm{\theta}_m + \alpha \nabla_{\bm{\theta}} \mathbb{E}_{\bm{\tau} \sim \mathrm{P}(\bm{\tau} \mid \pi)} \left[ J(\bm{\tau}) \right].
\label{eq:update_rule_general}
\end{equation}
This update allows the policy to transition from epoch $m$ to epoch $m+1$ at a learning rate $\alpha \in \mathbb{R}$.

In particular, we parameterize the policy using a deep neural network (DNN), modeling it as a Gaussian distribution over the control inputs:
\begin{equation} 
\bm{m}_t, \bm{\sigma}_t = \bm{f_{\mathrm{DNN}}} (\bm{s}_t, \bm{\Theta}), \, \bm{u}_t \sim \mathcal{N} (\bm{m}_t, \operatorname{diag}(\bm{\sigma}_t^2)),\label{eq:mean_xtd_policy} 
\end{equation} 
where $\bm{m}_t \in \mathbb{R}^{n_u}$ and $\bm{\sigma}_t \in \mathbb{R}^{n_u}$ denote the mean and standard deviation of the input distribution, respectively. The parameter vector $\bm{\Theta}$ represents the weights and biases of the DNN, thus $\bm{\theta}:=\bm{\Theta}$. The control input is then sampled from the resulting Gaussian distribution. Such a stochastic control policy allows the RL agent to naturally explore ({widening the distribution}) and exploit ({narrowing} the distribution) over epochs, gradually improving the expectation in \eqref{eq:rl_problem}. It is expected that deterministic or low-uncertainty systems will result in policies with very narrow distributions, leaning toward a deterministic input.

Upon applying the Policy Gradient Theorem \citep{NIPS1999_464d828b} and incorporating Eq. \eqref{eq:prob_dis_tau}, the gradient in Eq. \eqref{eq:update_rule_general} can be rewritten as:
\begin{equation}
    \nabla_{\bm{\theta}} \mathbb{E}_{\bm{\tau} \sim \mathrm{P}(\bm{\tau} \mid \pi)} \left[ J(\bm{\tau}) \right] = \int \mathrm{P}(\bm{{\tau}} \mid \bm{\theta}) \nabla_{\bm{\theta}} \log \mathrm{P}(\bm{\tau} \mid \bm{\theta}) J(\bm{\tau}) \, \mathrm{d}\bm{\tau} =\mathbb{E}_{\bm{\tau} \sim \mathrm{P}(\bm{\tau} \mid \pi)} \left[ J(\bm{\tau}) \nabla_{\bm{\theta}} \sum_{t=0}^{N_x-1} \log \pi(\bm{u}_t \mid \bm{s}_t, \bm{\theta}) \right].
\label{eq:policy_gradient_theorem}
\end{equation}

The otherwise intractable expectation in the previous equation is approximated using a Monte Carlo sampling over $N_\mathrm{MC}$ sampled trajectories (or episodes) within the epoch:
\begin{equation}
\nabla_{\bm{\theta}} \mathbb{E}_{\bm{\tau} \sim \mathrm{P}(\bm{\tau} \mid \pi)} \left[ J(\bm{\tau}) \right] \approx \frac{1}{N_\mathrm{MC}} \sum_{k=1}^{N_\mathrm{MC}} 
\bigg[ \frac{J\left( \bm{\tau}^{(k)} \right) - \Bar{J}_m\left( \bm{\tau} \right)}{\sigma_{J_{m}} + \epsilon_\mathrm{mach}} \nabla_{\bm{\theta}} \sum_{t=0}^{N_x-1} \log \left( \pi(\bm{u}_t^{(k)} \mid \bm{s}_t^{(k)}, \bm{\theta}) \right) \bigg],
\label{eq:grad_exp_2}
\end{equation}
where the return function is normalized by subtracting the mean return $\Bar{J}_m$ and dividing by the standard deviation of the return values $\sigma_{J_{m}}$ within the epoch. For numerical convenience, a small constant $\epsilon_\mathrm{mach}$ is added to the denominator to prevent division by zero, particularly in cases where the system and policy become deterministic.

\subsection{Domain randomization}
\label{subsec:domain_rand}
We implement domain randomization to generate policies that are robust to system uncertainties. Rather than mechanistically or explicitly modeling all sources of uncertainty, which can be particularly challenging in biotechnological system models, we define probability distributions from which disturbances are sampled. Doing so allows the computation of the state transition in Eq. \eqref{eq:dynamics_stoc} under uncertainty. These probability distributions are built based on domain knowledge or empirical data. By incorporating these uncertainties during training, each Monte Carlo trajectory experiences stochastic variations, allowing the policy to generalize across a wide range of possible stochastic dynamics.

Without loss of generality, in our case study, we consider uncertainties in both the initial conditions and key model kinetic parameters. Randomization of initial conditions can be useful to capture measurement errors and variability in growth media or inoculum conditions. Similarly, randomization of kinetic parameters can be useful to capture intrinsic stochastic intracellular phenomena, external process disturbances (e.g., temperature, pH, mixing variability), or wrong/oversimplified model assumptions. Specifically, domain randomization in each Monte Carlo episode $k$ is incorporated as follows:
\begin{subequations}
\begin{align}
       &\bm{x}_0^{(k)} = \bm{x}_0 + \bm{d}_x,\, \bm{d}_x \sim \mathcal{N} \big( \bm{0}, \operatorname{diag}(\sigma_x^2) \big) \label{eq:x0_rand}\\
       &\bm{\omega}^{(k)} = \bm{\omega} + \bm{d}_\omega, \, \bm{d}_\omega \sim \mathcal{N} \big( \bm{0}, \operatorname{diag}(\sigma_\omega^2) \big) \label{eq:par_rand}
\end{align}
\end{subequations}
where $\bm{d}_x$ and $\bm{d}_\omega$ are normally distributed zero-mean random variables of appropriate dimensions with predefined variances $\sigma_x^2$ and $\sigma_\omega^2$, respectively. The overall disturbance vector is then defined as $\bm{d}_t := [ \bm{d}_x^{\tran}, \bm{d}_\omega^{\tran}]^\tran$ (cf. Eq. \eqref{eq:dynamics_stoc}). Although Eqs. \eqref{eq:x0_rand}-\eqref{eq:par_rand} assume a Gaussian distribution, alternative distributions may be used based on prior knowledge of system uncertainties.

\section{{Biological systems}}
\label{sec:biological_system}
{Below, we outline the two biological systems that we consider as case studies to demonstrate our RL framework for robust dynamic metabolic control.} 

\subsection{{Engineered {\normalfont{\textit{E. coli}}} for fatty acid biosynthesis with ACC modulation}}
\label{subsec:fatty_acid_system}

\begin{figure} [htb!]
    \centering
    \includegraphics[scale=0.45]{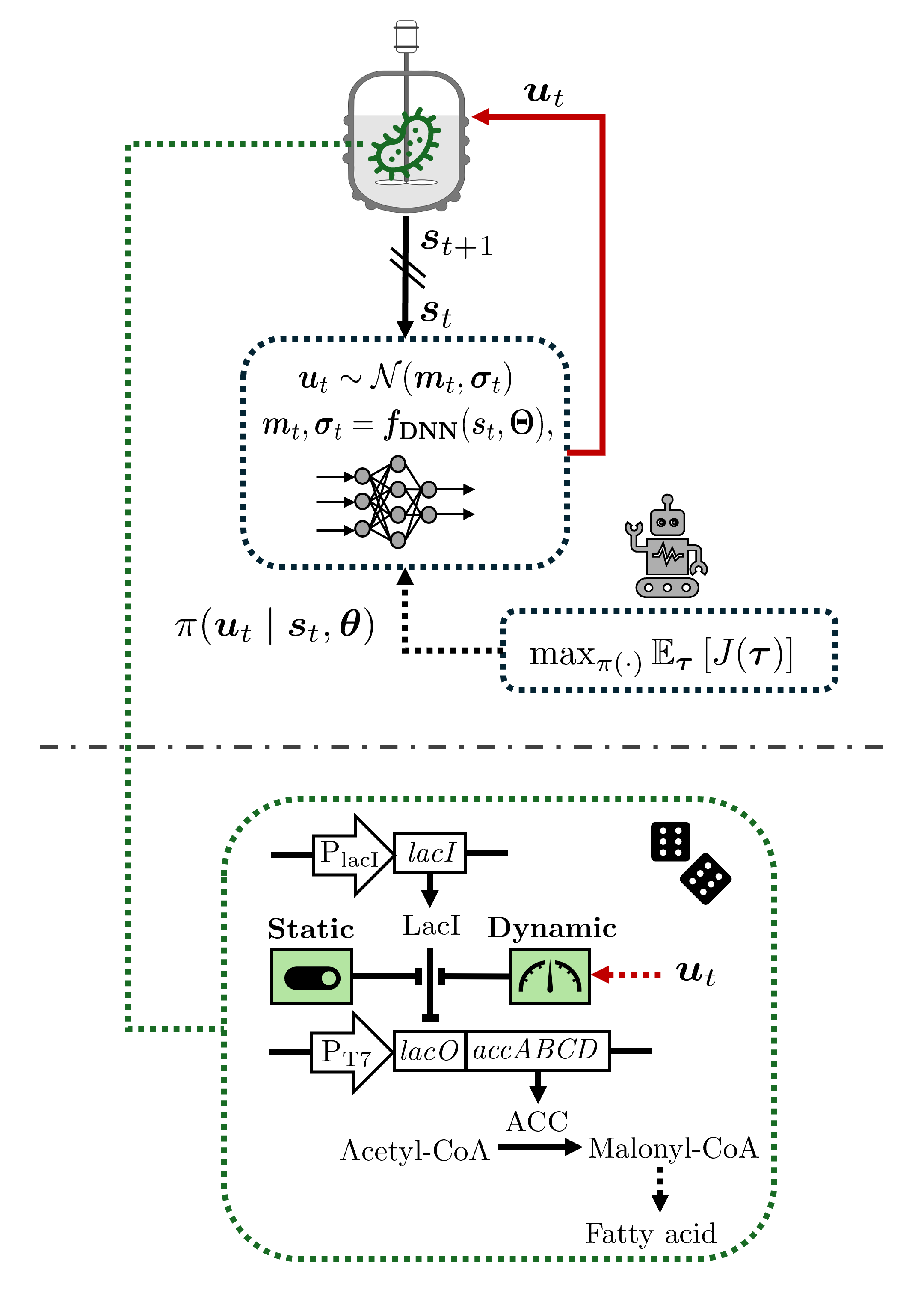}
    \caption{{Overview of the first case study. RL framework for robust dynamic metabolic control coupled to a fatty acid biosynthetic process with ACC modulation. $\bm{u}_t$: action/input at time $t$; $\bm{s}_t$: featurized system representation at time $t$; $\pi(\cdot)$: stochastic policy; $\bm{m}_t, \bm{\sigma}_t$: mean and standard deviation of Gaussian policy at time $t$; $\bm{\tau}$: joint trajectory of observed states, actions, and rewards; $J(\bm{\tau})$: return over $\bm{\tau}$; $\bm{\theta}$: policy parameters; $\bm{\Theta}$: deep neural network (DNN) parameters}.
    }
    \label{fig:cs_1}
\end{figure}

A diagram of {an engineered metabolic system for fatty acid biosynthesis by \textit{E. coli}, considered as our first case study,} is shown in Fig. \ref{fig:cs_1}. This system is motivated by the previous work of \cite{ohkubo_hybrid_2024}. LacI is a protein constitutively expressed by the cell under the regulation of the P\textsubscript{lacI} promoter. When LacI binds to the \textit{lacO} sequence, it blocks the expression of the enzyme ACC (encoded by \textit{accABCD}), regulated by the P\textsubscript{T7} promoter. ACC catalyzes the conversion of acetyl-CoA to malonyl-CoA, a key intermediate in the fatty acid biosynthesis pathway.

To simulate the bioprocess for fatty acid synthesis by \textit{E. coli} in batch fermentation regime, we consider the following dynamics for the concentrations of glucose $S \in \mathbb{R}$, {biomass} $X \in \mathbb{R}$, {manipulatable enzyme (in this case, ACC)} $E \in \mathbb{R}$, malonyl-CoA $M \in \mathbb{R}$, LacI $R \in \mathbb{R}$, and fatty acid $P \in \mathbb{R}$ \citep{ohkubo_hybrid_2024}:
\begin{subequations}
\begin{align}
    &\odv{S}{t} = -\mu \cdot {X^*}, \label{eq:ode_s_ohkubo} \\ 
    &\odv{{X^*}}{t} = \mu \cdot {X^*} - \mu_d \cdot {X^*}, \\
    &\odv{E}{t} = k_E \cdot \frac{K_{R_0}^{n_R}}{K_{R_0}^{n_R} + \left( \frac{R}{1 + \left(\frac{I}{K_I}\right)^{n_I}}\right)^{n_R}} - (d_E + \mu) \cdot E, \label{eq:ode_acc} \\
    &\odv{M}{t} = k_M \cdot E - k_P \cdot M - \mu \cdot M, \\
    &\odv{R}{t} = k_{R_1} - (d_R + \mu) \cdot R, \label{eq:ode_R} \\
    &\odv{{P^*}}{t} = k_P \cdot M \cdot {X^*} \cdot \left( \frac{S}{K_{S_P} + S} \right) \cdot (1 - T_P). \label{eq:ode_P} \\
    &X = H_X X^* \label{eq:x_meas} \\ 
    &P = H_P P^* \label{eq:p_meas} \\
    &{S(t_0) = 1-X^*(t_0)}, {X^*}(t_0) = {X_0^*}, E(t_0) = E_0, M(t_0) = M_0, R(t_0) = R_0, {P^*}(t_0) = {P_0^*}.
\end{align}
\end{subequations}
All the dynamic states involved in the differential equations of this model {(cf. Eqs. \eqref{eq:ode_s_ohkubo}–\eqref{eq:ode_P}) are normalized dimensionless variables, consistent with the original model of \cite{ohkubo_hybrid_2024}. For clarity, the notation $X^*$ and $P^*$ is used specifically for biomass and product to distinguish their normalized dimensionless state variables from the corresponding experimental measurements. Following Eqs. \eqref{eq:x_meas}-\eqref{eq:p_meas}, $H_X$ converts the dimensionless biomass concentration $X^*$ to relative cell density calculated based on the optical density (OD\textsubscript{600}), while $H_P$ converts the dimensionless product concentration $P^*$ to $\mathrm{g/L}$.} The growth rate function $\mu$ and the toxic effects of ACC expression on biomass and product formation, represented by $T_X$ and $T_P$, are governed by \citep{ohkubo_hybrid_2024}:
\begin{subequations}
\begin{align}
    \mu &= k_X \cdot S \cdot \left(1 - T_X\right), \label{eq:mu_c1}\\
    T_X &= T_{X_{\text{max}}} \frac{E^{n_{T_X}}}{K_{T_X}^{n_{T_X}} + E^{n_{T_X}}}, \\
    T_P &= \begin{cases}
        0, & E < E_{\text{tox}} \\
        \frac{(E - E_{\text{tox}})^{n_{T_P}}}{K_{T_P}^{n_{T_P}} + (E - E_{\text{tox}})^{n_{T_P}}}, & E \geq E_{\text{tox}}. \label{eq:T_P}
    \end{cases}
\end{align}
\end{subequations}
{Here, $k_E$, $\mu_d$, $K_{R_0}$, $n_R$, $K_I$, $n_I$, $d_E$, $k_M$, $k_P$, $k_{R_1}$, $d_R$, $K_{S_P}$, $k_X$, $T_{X_{\text{max}}}$, $K_{T_X}$, $n_{T_X}$, $E_{\text{tox}}$, $K_{T_P}$, and $n_{T_P}$ are constant model parameters governing gene regulation, enzyme and process kinetics, and inhibition thresholds. For more details on the modeling assumptions and derivation, the reader is referred to \citep{ohkubo_hybrid_2024}.}

$T_X$ and $T_P$ reduce the growth rate (cf. Eq. \eqref{eq:mu_c1}) and the product formation rate (cf. Eq. \eqref{eq:ode_P}), respectively. $T_X$ follows Hill-type activation kinetics, while $T_P$ is a piecewise function that follows Hill-activation kinetics once a certain enzyme expression threshold $E_{\text{tox}}$ is reached. Therefore, it is clear that the overall system dynamics are rate-limited by both the substrate glucose and the intracellular ACC concentration.

The rate of {ACC} expression (cf. Eq. \eqref{eq:ode_acc}) is influenced by the control input (inducer) denoted as $I \in \mathbb{R}$ and is assumed to come from an external source. Hereafter, for consistency of notation with Section \ref{sec:RL_domain_rand}, we refer to {this} inducer as ${u}$. {Thus, in the RL context, $\bm{u}_t := u_t \in \mathbb{R}$}. Depending on the metabolic control strategy (i.e., static or dynamic), $u$ will either remain constant throughout the process in the static approach or vary over time in the dynamic approach. In practice, the static approach can involve the use of chemical inducers, such as IPTG, which is added to the cultivation medium at a specific concentration \citep{ohkubo_hybrid_2024}. When IPTG binds to LacI, it causes LacI to dissociate from \textit{lacO}, resulting in the transcription of target genes. In contrast, the dynamic approach involves bidirectionally tunable inputs, which can be achieved, e.g., with the OptoLacI system \citep{liu_optolaci_2024}, where LacI is engineered to respond to light instead of IPTG. For ease of comparison, we assume that the model’s parameter values hold regardless of the metabolic control strategy applied (static or dynamic).

\subsection{{Engineered {\normalfont{\textit{E. coli}}} for lactate biosynthesis with ATPase modulation}}
\label{subsec:lactate_system}
\begin{figure} [htb!]
    \centering
    \includegraphics[scale=0.45]{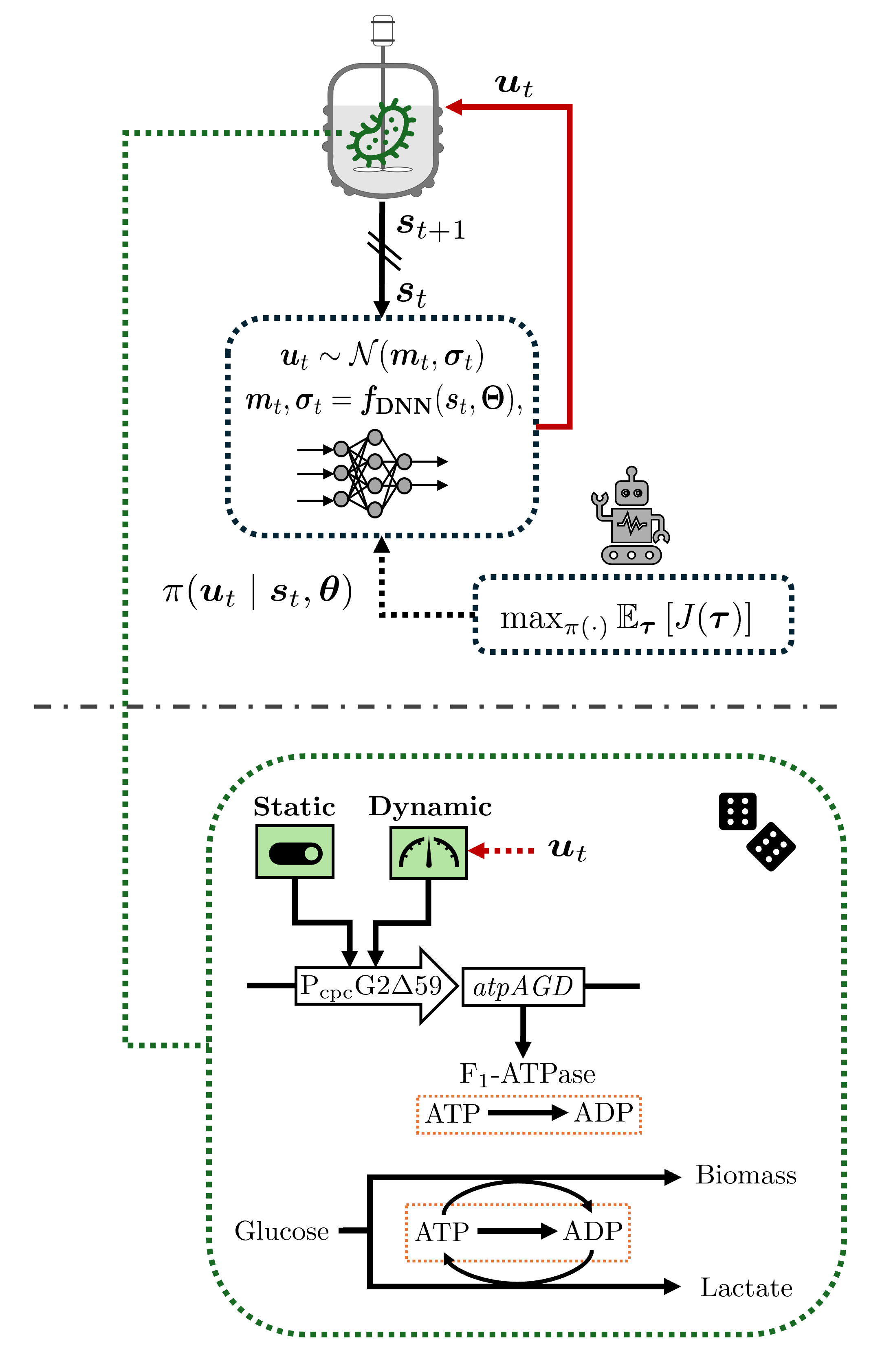}
    \caption{{Overview of the second case study. RL framework for robust dynamic metabolic control coupled to a lactate biosynthetic process with ATPase modulation. $\bm{u}_t$: action/input at time $t$; $\bm{s}_t$: featurized system representation at time $t$; $\pi(\cdot)$: stochastic policy; $\bm{m}_t, \bm{\sigma}_t$: mean and standard deviation of Gaussian policy at time $t$; $\bm{\tau}$: joint trajectory of observed states, actions, and rewards; $J(\bm{\tau})$: return over $\bm{\tau}$; $\bm{\theta}$: policy parameters; $\bm{\Theta}$: deep neural network (DNN) parameters}.
    }
    \label{fig:cs_2}
\end{figure}

{A diagram of an engineered metabolic system for lactate biosynthesis by \textit{E. coli}, considered as our second case study, is shown in Fig. \ref{fig:cs_2}. This system is motivated by the previous work of \cite{espinel-rios_experimentally_2024}. The engineered strain has deletions in the ethanol and acetate pathways ($\Delta$\textit{adhE}, $\Delta$\textit{ackA-pta}) and produces lactate as the main fermentation product under anaerobic conditions. It also carries a green-light-inducible system (CcaS/CcaR) that regulates the expression of the ATPase F\textsubscript{1}-subunit (\textit{atpAGD}) under the promoter PcpcG2Δ59. ATPase catalyzes the hydrolysis of ATP into ADP. Here, lactate biosynthesis is coupled to net ATP production through the redox cofactor regeneration required during glycolysis. Therefore, increasing ATPase expression raises ATP turnover (i.e., \textit{ATP wasting}), which in turn increases substrate uptake and lactate fluxes as a compensatory mechanism, yet at the expense of reduced cell growth.}

{To simulate the bioprocess for lactate synthesis by \textit{E. coli} in batch fermentation regime, we consider the following dynamics for the concentrations of glucose, biomass, {manipulatable enzyme (in this case, ATPase)}, and lactate $L \in \mathbb{R}$ \citep{espinel-rios_experimentally_2024}:}
{\begin{subequations} 
\begin{align} 
&\odv{S}{t} = -q_S \cdot X, \label{eq:ode_glc} \\ 
&\odv{X}{t} = \mu \cdot X, \label{eq:ode_bio} \\
&\odv{L}{t} = q_L \cdot X,\label{eq:ode_lac} \\
&\odv{E}{t} = q_E - k_d \cdot E,\label{eq:ode_E} \\
&S(t_0)=S_0, X(t_0)=X_0, L(t_0)=L_0, E(t_0)=E_0.
\end{align}
\end{subequations}}

{Note that, as in the model for fatty acid biosynthesis, we also denote the glucose, biomass, and manipulatable enzyme states as $S$, $X$, and $E$, respectively. However, here $X$ and $S$ are represented in $\mathrm{g/L}$. $E$ is represented in arbitrary (\textit{virtual}) units per gram of biomass ($\mathrm{VU/g}$), not measured during model development in \citep{espinel-rios_experimentally_2024}, serving as an abstract representation of ATPase accumulation. The growth rate function $\mu$, the substrate uptake rate $q_S$, the lactate synthesis rate $q_L$, and the light-dependent expression rate of ATPase are governed by (adapted from \cite{espinel-rios_experimentally_2024}):} 
{\begin{subequations} 
\begin{align} 
&q_S = q_{S_\mathrm{max}}\left( \frac{S}{S+k_S}\right) \left( 1 + \frac{E^{n_1}}{E^{n_1}+k_\mathrm{SV}^{n_1}} \right),\label{eq:q_S} \\ 
&\mu = Y_\mathrm{XS}\left( q_S-m_S \right) \left( 1 - \frac{E^{n_2}}{E^{n_2}+k_\mathrm{XV}^{n_2}} \right),\label{eq:mu_c2} \\
&q_L = \left(Y_\mathrm{LX} \cdot \mu + m_L\frac{S}{S+k_\mathrm{LS}}\right) \left( 1 + \frac{E^{n_3}}{E^{n_3}+k_\mathrm{LV}^{n_3}} \right),\label{eq:q_L} \\
&q_E = q_{E_0} + q_{E_\mathrm{max}} \frac{l^{n_4}}{l^{n_4}+k_l^{n_4}},\label{eq:q_E}
\end{align}
\end{subequations}}
{Here, $k_\mathrm{XV}$, $k_S$, $k_\mathrm{SV}$, $k_\mathrm{LV}$, $m_S$, $m_L$, $k_\mathrm{LS}$, $n_1$, $n_2$, $n_3$, $q_{S_\mathrm{max}}$, $Y_\mathrm{XS}$, $Y_\mathrm{LX}$, $q_{E_0}$, $q_{E_\mathrm{max}}$, $n_4$, $k_l$, $k_d$ are constant model parameters governing gene regulation and process kinetics. For more details on the modeling assumptions and derivation, the reader is referred to \citep{espinel-rios_experimentally_2024}.}

{Overall, following an increase in ATPase expression, and consequently intracellular ATP turnover, the model captures the expected increase in substrate uptake and lactate synthesis rates, alongside a decrease in growth rate. As in the fatty acid biosynthesis case study, the overall system dynamics here are also rate-limited by both the substrate glucose and the manipulatable enzyme concentration.}

{The rate of ATPase expression (cf. Eq. \eqref{eq:q_E}) is influenced by the control input (inducer) denoted as $l \in \mathbb{R}$ which represents green light photon flux density in \textmu$\mathrm{mol \, m^{-2} \, s^{-1}}$. Hereafter, for consistency of notation with Section \ref{sec:RL_domain_rand}, we refer to this inducer as ${u}$; $\bm{u}_t := u_t \in \mathbb{R}$.}

\section{{Framework demonstration}}
\label{sec:control_results}
{To assess the effectiveness of our RL-based framework for deriving robust dynamic metabolic control policies, we apply it to the two case studies outlined in the previous section.}

\subsection{{RL-driven dynamic metabolic control in fatty acid biosynthesis with ACC modulation}}
{We first focus on the fatty acid biosynthetic} system described in Section  \ref{subsec:fatty_acid_system} under varying levels of uncertainty. Specifically, we introduce uncertainty levels of 0 \%, 10 \%, 15 \%, 20 \%, and 25 \% in the initial conditions and in the parameters $k_E$ and $k_{R_1}$ (cf. Eqs. \eqref{eq:ode_acc} and \eqref{eq:ode_R}), which regulate input-dependent ACC expression and basal LacI expression, respectively. The randomization follows the procedure outlined in Section \ref{subsec:domain_rand}. {Values for nominal parameters and initial conditions for this {system were} taken from \citep{ohkubo_hybrid_2024} {(see Table \ref{tab:cs1_params_ic} for details)}.}

\begin{table}[htb!]
\centering
\caption{{Fatty acid biosynthesis case study with ACC modulation: nominal parameters and initial conditions.}}
\label{tab:cs1_params_ic}
\begin{tabular}{lcl}
\hline
\textbf{Symbol} & \textbf{Value} & \textbf{Unit} \\
\hline
$k_X$                 & 0.4639   & $\mathrm{1/h}$ \\
$k_E$                 & 0.6088   & $\mathrm{1/h}$ \\
$k_M$                 & 0.4314   & $\mathrm{1/h}$ \\
$k_P$                 & 0.4314   & $\mathrm{1/h}$ \\
$k_{R_1}$             & 17.77    & $\mathrm{1/h}$ \\
$\mu_d$               & 0.00763  & $\mathrm{1/h}$ \\
$T_{X_{\text{max}}}$  & 0.5081   & -- \\
$E_{\text{tox}}$      & 1.0      & -- \\
$d_E$                 & 0.1131   & $\mathrm{1/h}$ \\
$d_R$                 & 1.386    & $\mathrm{1/h}$ \\
$K_{T_X}$             & 0.4587   & -- \\
$K_{T_P}$             & 0.3445   & -- \\
$K_{R_0}$             & 1.0      & -- \\
$K_I$                 & 17.61    & \textmu$\mathrm{M}$ \\
$K_{S_P}$             & 0.01397  & -- \\
$n_{T_X}$             & 2.798   & -- \\
$n_{T_P}$             & 1.137    & -- \\
$n_R$                 & 0.5576   & -- \\
$n_I$                 & 1.034    & -- \\
$H_X$                 & 1.688      & -- \\
$H_P$                 & 0.4843      & $\mathrm{g/L}$ \\
\hline
$X_0^*$               & 0.1107   & -- \\
$E_0$                 & 0.0      & -- \\
$M_0$                 & 0.0      & -- \\
$R_0$                 & 0.002    & -- \\
$P_0^*$               & 0.0      & -- \\
$S_0$                 & $1 - X_0^*$ & -- \\
\hline
\end{tabular}
\end{table}

The scenario with 0 \% uncertainty, representing purely deterministic dynamics, serves as a reference for an ideally behaved system and initially tests the ability of our method to converge to an optimal result. For benchmarking, we compare our dynamic metabolic control approach against a static metabolic control strategy from a previous study, where an optimal constant input (IPTG concentration, \textmu M) of $u \approx 40$ was identified {for maximizing product titer}  \citep{ohkubo_hybrid_2024}. {This static control strategy serves as a baseline for evaluating the benefits of applying dynamic metabolic control.} {In our first case study, we selected the final fatty acid titer as the return function to maximize (cf. RL problem in \eqref{eq:rl_problem}); thus:}
\begin{equation} 
{\mathbb{E}{\left[ J(\cdot) \right]} = \mathbb{E}{\left[ P(t_{N_x}) \right]}.} \label{eq:J_c1} 
\end{equation}
{In a batch process, this is equivalent to maximizing the product volumetric productivity within the given time frame.}

{\textit{Remark on RL training}}. {The design and hyperparameters of the RL policy were set based on previous studies rendering good convergence \citep{petsagkourakis_reinforcement_2020,espinel-rios_enhancing_2024,espinel-rios_reinforcement_2025}. Only the learning rate was selected through grid search based on faster convergence. In summary, a fully connected feedforward neural network} with four hidden layers, each containing 20 nodes and employing LeakyReLU activation functions, was used to parameterize the policy. The RL policy was trained in PyTorch \citep{pytorch} over 350 epochs, each consisting of 500 episodes {or trajectories}, using a learning rate of $\alpha = 0.0075$. {The policy’s output mean was constrained to the interval $[u_\mathrm{lb},\,u_\mathrm{ub}]$. The policy’s output standard deviation was constrained to at most 25 \% of the input range, i.e., $0.25\,(u_\mathrm{ub}-u_\mathrm{lb})$. In the fatty acid case study, the input range was defined as $[0,1000]$, consistent with the work of \citep{ohkubo_hybrid_2024}.} Stepwise constant inputs, applied at 1-h intervals, were used over a total process duration of 25 h. The feature vector $\bm{s}_t$ consisted of the two most recent state-input pairs and a process time embedding, normalized to the range $[-1,1]$. {Furthermore, we implemented early stopping such that the training would stop after 50 consecutive epochs without improvement in the mean return}. Full state observability was assumed, as the idea of our proposed methodology is to use the dynamic mathematical model as a surrogate environment, which in principle is fully observable. 

\subsubsection{Deterministic dynamics}
\begin{figure}[htb!]
    \begin{center}
        \subfigure[\,\textbf{Return function}]{\includegraphics[scale=0.38, trim={0 10 0 10}, clip]{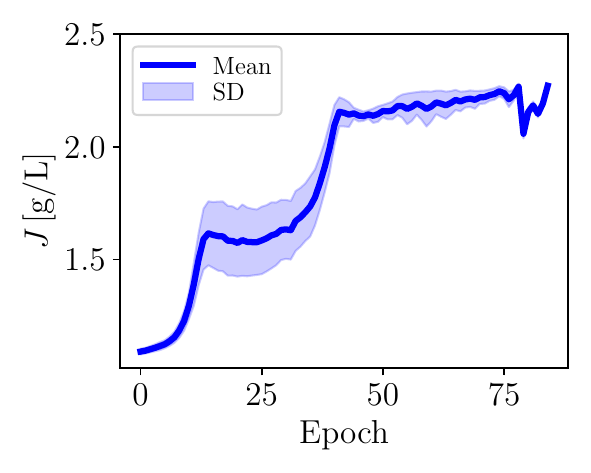}} 
        \subfigure[\,\textbf{Input}]{\includegraphics[scale=0.38, trim={0 10 0 10}, clip]{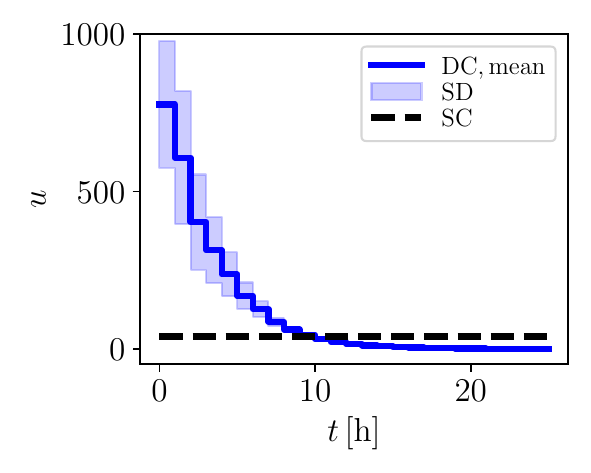}} 
        \subfigure[\,\textbf{Cell density}]{\includegraphics[scale=0.38, trim={0 10 0 10}, clip]{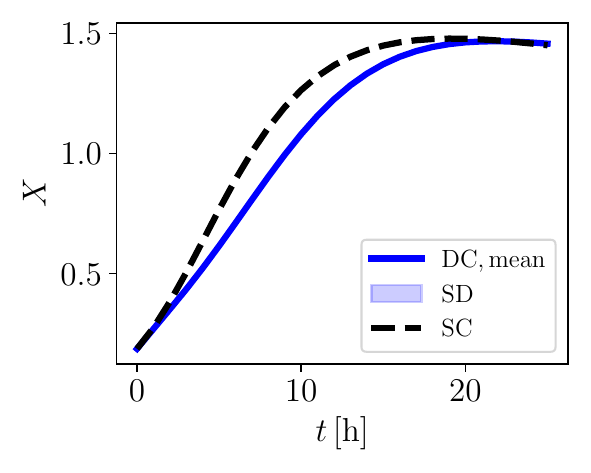}} 
        \subfigure[\,\textbf{Glucose}]{\includegraphics[scale=0.38, trim={0 10 0 10}, clip]{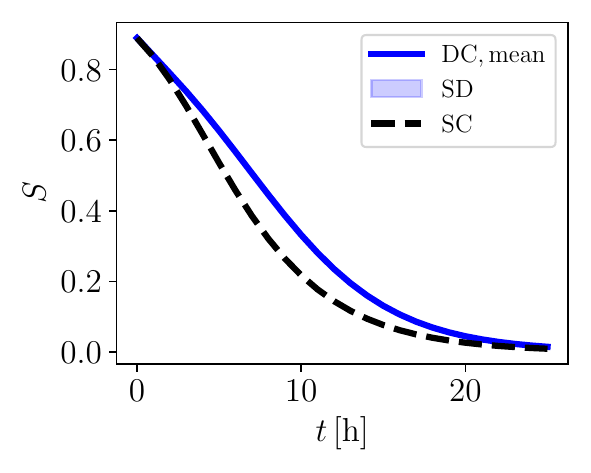}} 
        \subfigure[\,\textbf{LacI}]{\includegraphics[scale=0.38, trim={0 10 0 10}, clip]{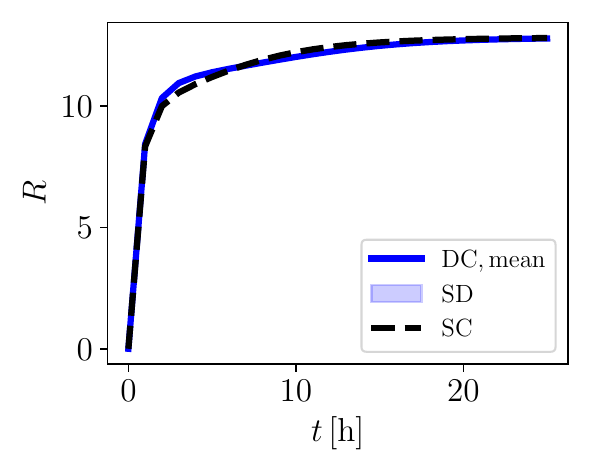}} 
        \subfigure[\,\textbf{Malonyl-CoA}]{\includegraphics[scale=0.38, trim={0 10 0 10}, clip]{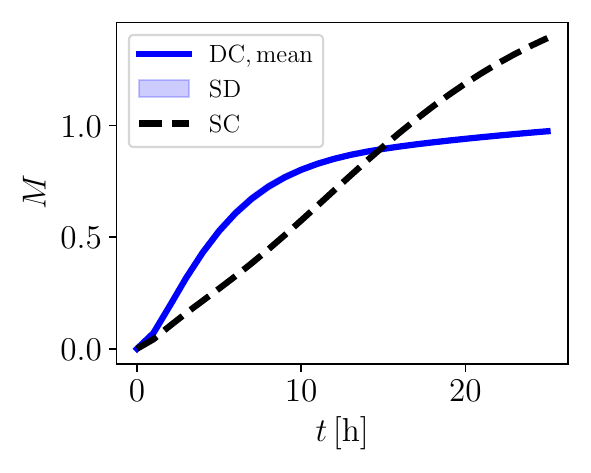}}
        \subfigure[\,\textbf{ACC}]{\includegraphics[scale=0.38, trim={0 10 0 10}, clip]{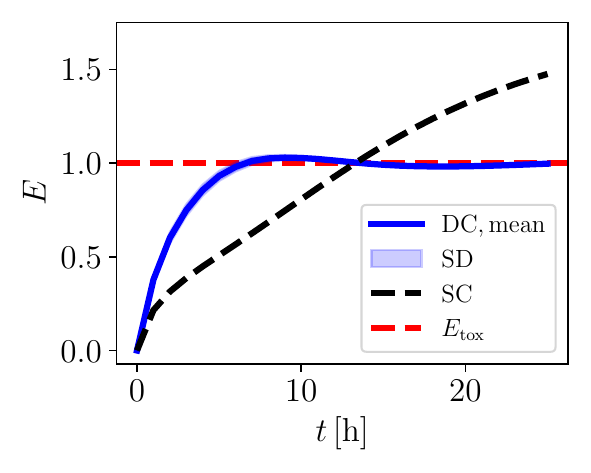}} 
        \subfigure[\,\textbf{Fatty acid}]{\includegraphics[scale=0.38, trim={0 10 0 10}, clip]{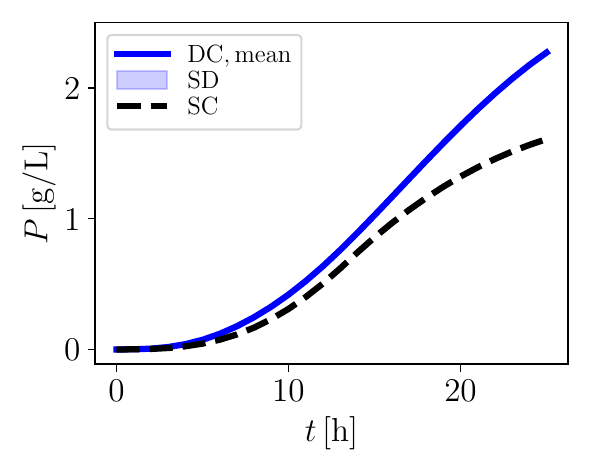}} 
        \caption{Metabolic control results under ideal conditions (i.e., no system uncertainties) {for the fatty acid biosynthesis case study with ACC modulation}. (a) Evolution of the return function over epochs, up to the epoch with the highest mean value (selected control policy). The corresponding (b) input trajectory and (c)-(h) dynamic state trajectories associated with the selected control policy are also shown. The RL-derived dynamic control scenario (DC) is benchmarked against the static control scenario (SC). {Uncertainty bands correspond to 500 episodes or trajectories.} SD: standard deviation. {$J$: return, $u$: control input (inducer); $X$: biomass; $S$: glucose; $R$: LacI; $M$: malonyl-CoA; $E$: manipulatable enzyme (ACC); $P$: fatty acid.}}
        \label{fig:0_unc}
    \end{center}
\end{figure}

Fig. \ref{fig:0_unc} shows the performance of the {RL-driven} dynamic metabolic control {strategy} under ideal deterministic conditions (i.e., 0 \% uncertainty). As expected, the return function initially exhibits a wider standard deviation across epochs while the agent explores different policies (cf. e.g. the region around epoch 25). As training progresses and the return function converges, the standard deviation decreases, indicating a shift toward exploitation mode. This demonstrates the natural balance between exploration and exploitation inherent to our RL {framework} based policy gradients, eliminating the need for heuristic exploration strategies.

The {RL-driven} dynamic metabolic control strategy follows a gradually decreasing input trend, enabling precise modulation of ACC enzyme expression. {This enables rapid accumulation of ACC, extending} the duration of active fatty acid production while carefully avoiding the toxicity threshold (cf. $E_{\text{tox}}$ in Eq. \eqref{eq:T_P}). In contrast, the static control scenario leads to a slower ACC accumulation rate, delaying but ultimately failing to prevent the system from surpassing the toxicity threshold due to the irreversible nature of static induction. For other system states, LacI accumulates steadily in both static and dynamic control scenarios, as expected due to its constitutive expression. Malonyl-CoA reaches higher levels under static control because increased ACC toxicity slows down fatty acid production, reducing malonyl-CoA conversion efficiency and leading to its accumulation. 

{Overall,} the final fatty acid titer increases by approximately 41 \% under dynamic control compared to static control (cf. Table \ref{tab:summary_results}). Thus, the RL-based {dynamic metabolic} control policy more effectively regulates ACC expression dynamics, {optimally managing} enzyme induction and its toxic effects on both cell growth and fatty acid biosynthesis.

\begin{table}[h!]
\begin{center}
\caption{{Final fatty acid titers under static and dynamic control policies across different uncertainty levels in the fatty acid biosynthesis case study with ACC modulation. {Prediction uncertainty corresponds to 500 episodes or trajectories.}}} \label{tab:summary_results}
\begin{tabular}{|c|c|c|c|}
    \hline
    Unc. [\%] &  SC [$\mathrm{g/L}$] & DC [$\mathrm{g/L}$] & Imp. [\%]\\
    \hline
    0  &  1.61 $\pm$ 0.00 & 2.27	$\pm$ 0.00 & 41 \% \\
    \hline
    10  &  1.57 $\pm$ 0.09 & 2.18 $\pm$ 0.20 & 39 \% \\
    \hline
    15  &  1.54 $\pm$ 0.13 & 2.07 $\pm$ 0.31 & 35 \% \\
    \hline
    20 &  1.46 $\pm$ 0.22 & 2.00 $\pm$ 0.39 & 37 \% \\
    \hline
    25 &  1.38 $\pm$ 0.30 & 1.93 $\pm$ 0.45 & 40 \% \\
    \hline
\end{tabular}\\
\smallskip\noindent
SC: static control. DC: dynamic control. Imp.: improvement. Unc.: uncertainty level.
\end{center}
\end{table}

\subsubsection{Policy robustness via domain randomization}
\begin{figure*}[htb!]
    \begin{center}
        \subfigure[\,\textbf{10 \% uncertainty}]{\includegraphics[scale=0.38, trim={0 10 0 10}, clip]{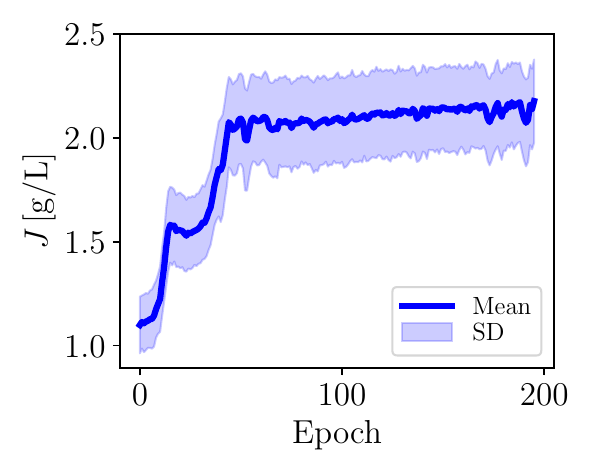}} 
        \subfigure[\,\textbf{15 \% uncertainty}]{\includegraphics[scale=0.38, trim={0 10 0 10}, clip]{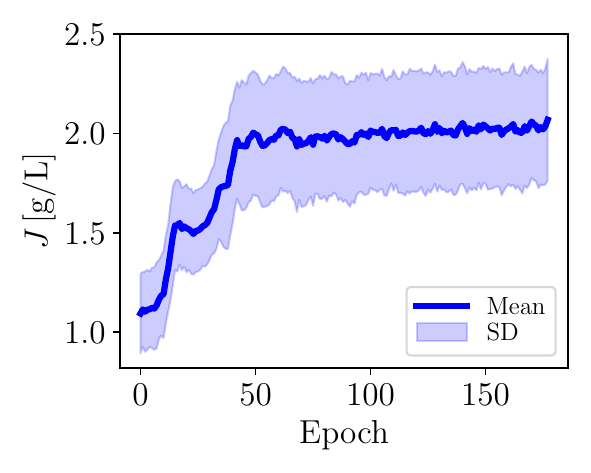}} 
        \subfigure[\,\textbf{20 \% uncertainty}]{\includegraphics[scale=0.38, trim={0 10 0 10}, clip]{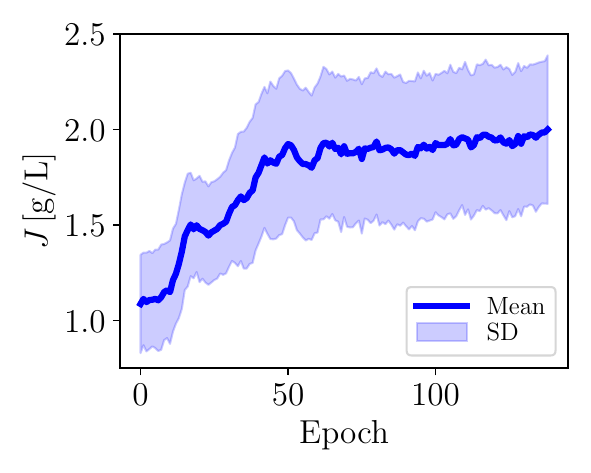}} 
        \subfigure[\,\textbf{25 \% uncertainty}]{\includegraphics[scale=0.38, trim={0 10 0 10}, clip]{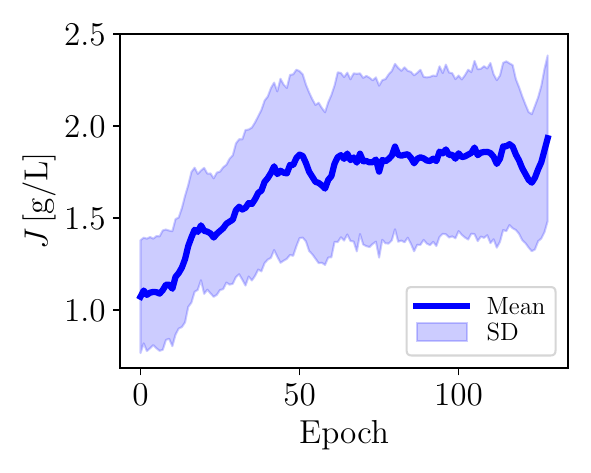}}

        \subfigure{\includegraphics[scale=0.38, trim={0 10 0 10}, clip]{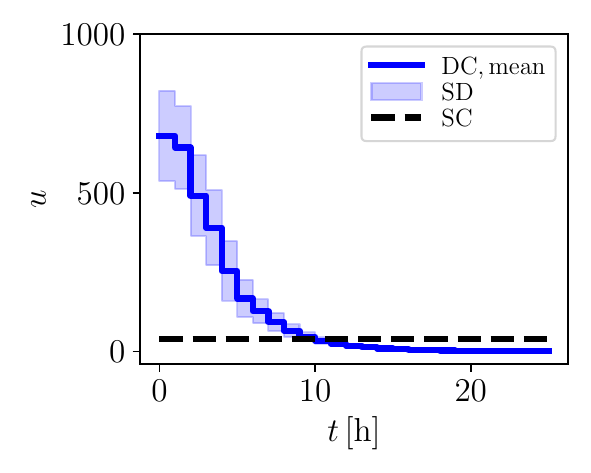}} 
        \subfigure{\includegraphics[scale=0.38, trim={0 10 0 10}, clip]{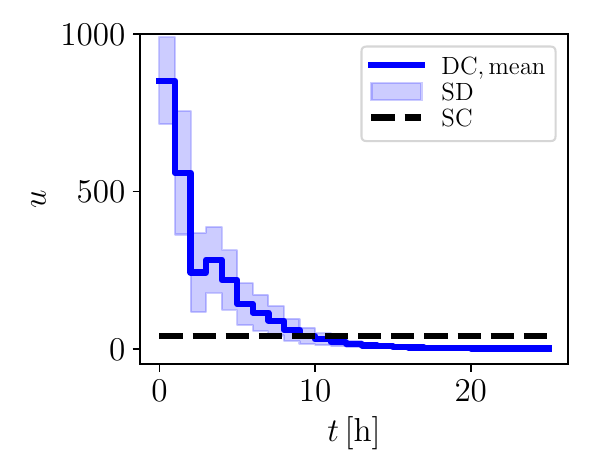}} 
        \subfigure{\includegraphics[scale=0.38, trim={0 10 0 10}, clip]{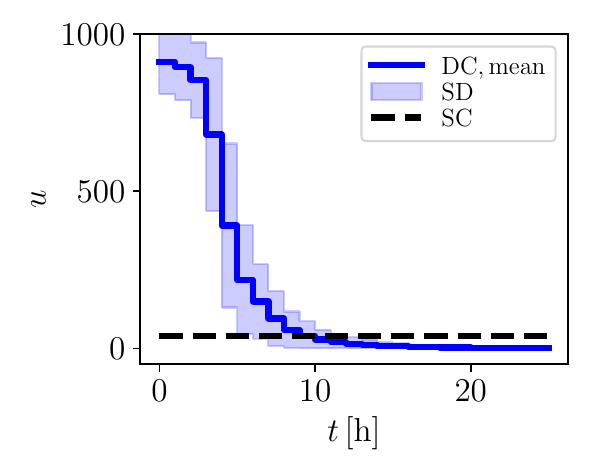}} 
        \subfigure{\includegraphics[scale=0.38, trim={0 10 0 10}, clip]{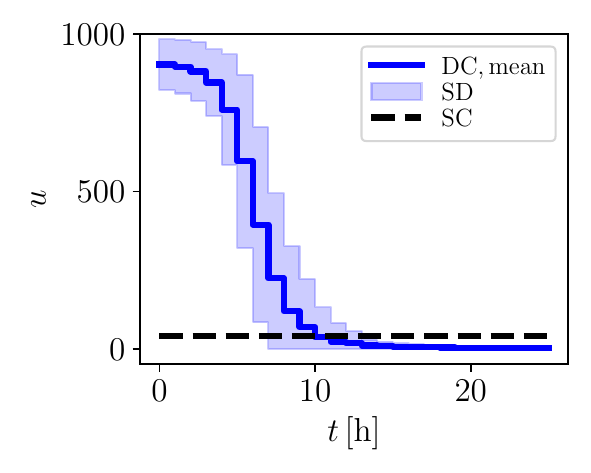}}

        \subfigure{\includegraphics[scale=0.38, trim={0 10 0 10}, clip]{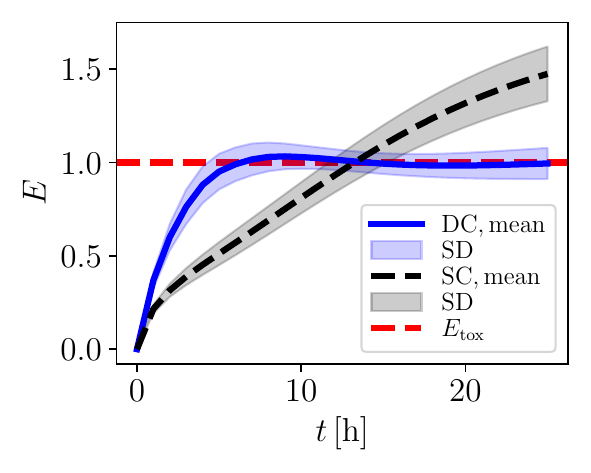}} 
        \subfigure{\includegraphics[scale=0.38, trim={0 10 0 10}, clip]{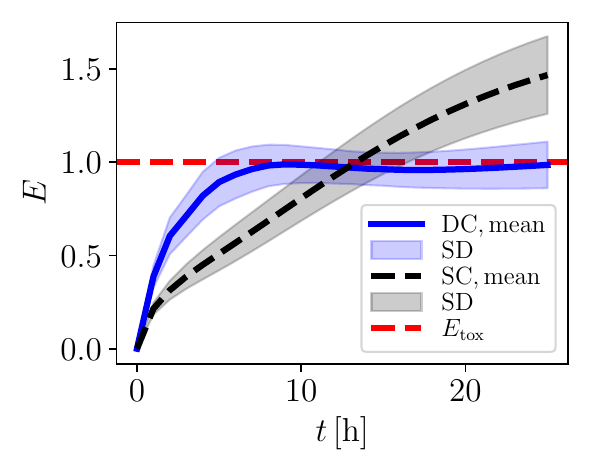}} 
        \subfigure{\includegraphics[scale=0.38, trim={0 10 0 10}, clip]{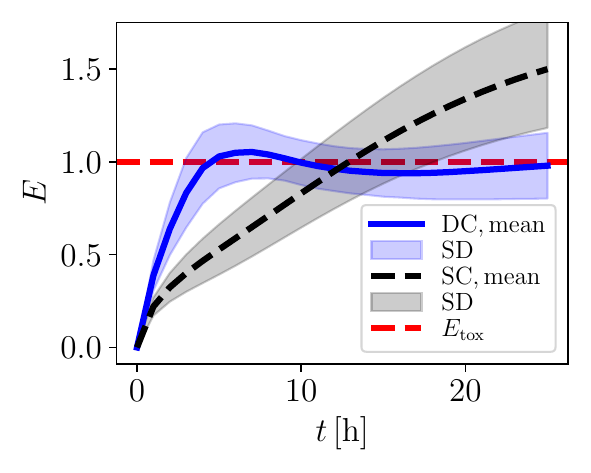}} 
        \subfigure{\includegraphics[scale=0.38, trim={0 10 0 10}, clip]{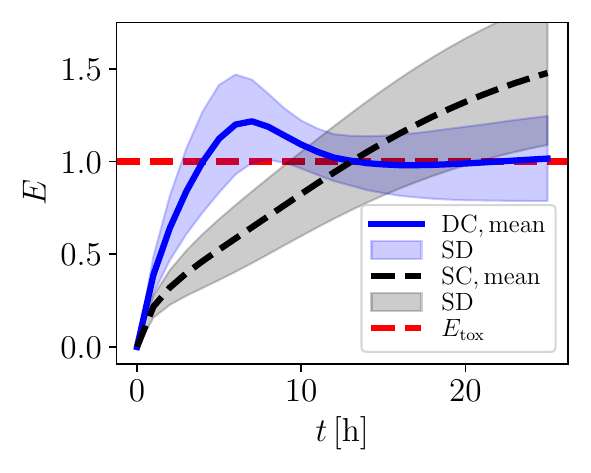}}

        \subfigure{\includegraphics[scale=0.38, trim={0 10 0 10}, clip]{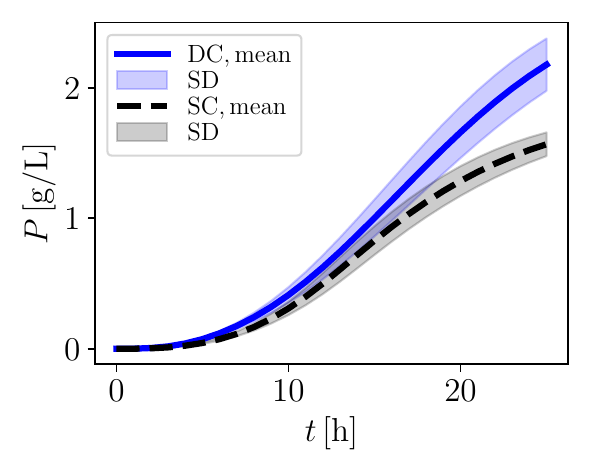}} 
        \subfigure{\includegraphics[scale=0.38, trim={0 10 0 10}, clip]{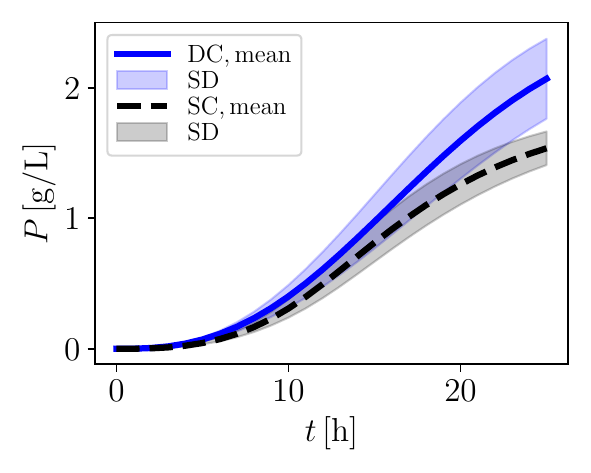}} 
        \subfigure{\includegraphics[scale=0.38, trim={0 10 0 10}, clip]{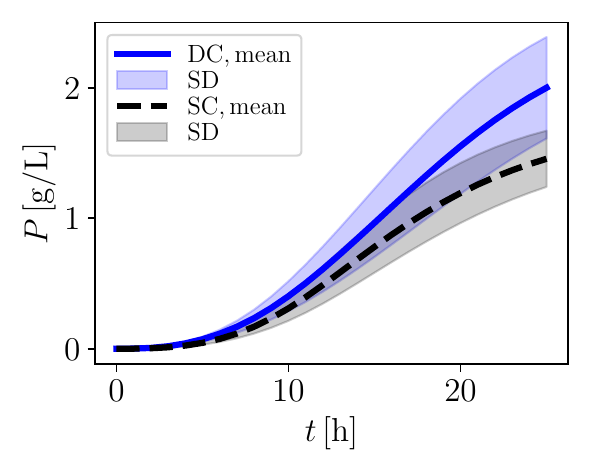}} 
        \subfigure{\includegraphics[scale=0.38, trim={0 10 0 10}, clip]{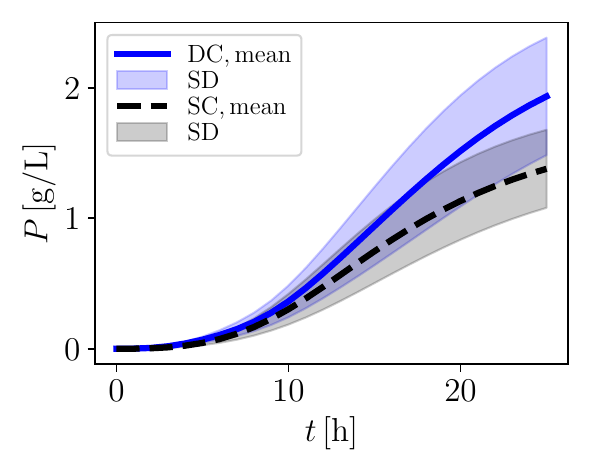}}
        
        \caption{Control policies robust against system uncertainty {for the fatty acid biosynthesis case study with ACC modulation}, considering (a) 10 \%, (b) 15 \%, (c) 20 \%, and (d) 25 \% uncertainty in the initial conditions and key kinetic parameters affecting the expression of LacI and ACC. The RL-derived dynamic control scenario (DC) is benchmarked against the static control scenario (SC). The return function is presented up to the epoch with the highest mean return value, matching the chosen policy. Selected dynamic state trajectories correspond to the latter policy. {Uncertainty bands correspond to 500 episodes or trajectories}. SD: standard deviation. {$J$: return, $u$: control input (inducer); $E$: manipulatable enzyme (ACC); $P$: fatty acid.}}
        \label{fig:fig_unc}
    \end{center}    
\end{figure*}

The performance of the {RL-driven} dynamic metabolic control strategy under varying levels of uncertainty is shown in Fig. \ref{fig:fig_unc}. As expected, higher uncertainty levels lead to greater standard deviations in both the return function evolution over epochs and the dynamic states for the best-performing epoch, reflecting the increased stochasticity in the bioprocess dynamics. Despite these variations, all dynamic control scenarios under uncertainty maintain a gradually decreasing mean input trend, consistent with the deterministic case.

While the optimized trajectories exhibit larger standard deviations, the RL-derived policies successfully regulate the \textit{mean} ACC concentration, keeping it below the toxicity threshold for fatty acid biosynthesis. However, at higher uncertainty levels (e.g., 20 \% and 25 \%), a slight transient overshoot is observed, yet the controller rapidly restores the mean ACC concentration to a non-toxic level. Notably, the RL-derived policy implicitly identifies the toxicity threshold, despite it being \textit{agnostic} to this information. That is, the toxicity threshold was not incorporated as a \textit{constraint} during training. Instead, this insight emerges naturally through exploration with the surrogate environment.

Across all uncertainty scenarios, the dynamic metabolic control strategy consistently achieves mean fatty acid titer improvements of 35–40 \% relative to the static control approach \textit{under the same uncertainty conditions} (cf. Table \ref{tab:summary_results}). Despite this consistently significant performance gain, mean fatty acid titers exhibit a slight decline as uncertainty levels increase. This trend is expected, as higher uncertainty inevitably reduces overall system efficiency and robustness. For instance, under the highest uncertainty scenario (25 \%), dynamic metabolic control performance decreases by approximately 15 \% relative to the deterministic case. However, this does not undermine the value of our RL approach; rather, it demonstrates its ability to work effectively even under highly variable system conditions.

\subsection{{RL-driven dynamic metabolic control in lactate biosynthesis with ATPase modulation}}
{To assess the generalizability of our RL-based framework for dynamic metabolic control, we further consider the lactate biosynthetic system described in Section \ref{subsec:lactate_system} under varying levels of uncertainty. Specifically, we introduce uncertainty levels of 0 \%, 5 \%, 10 \%, 12.5 \%, and 15 \% in the initial conditions and in the parameter $q_{E_\mathrm{max}}$ (cf. Eq. \eqref{eq:ode_E}), which governs the maximum rate of input-dependent ATPase expression. The randomization follows the procedure outlined in Section \ref{subsec:domain_rand}. As with the fatty acid biosynthesis case study, the scenario with 0 \% uncertainty represents the deterministic dynamics and initially tests the ability of our method to converge to an optimal result. Values for all nominal parameters were taken from \citep{espinel-rios_experimentally_2024}, except for $k_\mathrm{LS}$, which was set to $1 \times 10^{-10}\, \mathrm{g/L}$. Note that the latter parameter is not part of the original model in \citep{espinel-rios_experimentally_2024}, yet we introduced it to prevent production of lactate under substrate starvation conditions. {See Table \ref{tab:cs2_params_ic} for details on parameter values and initial conditions.}}

\begin{table}[htb!]
\centering
\caption{{Lactate biosynthesis case study with ATPase modulation: nominal parameters and initial conditions.}}
\label{tab:cs2_params_ic}
\begin{tabular}{lcl}
\hline
\textbf{Symbol} & \textbf{Value} & \textbf{Unit} \\
\hline
$q_{S_{\mathrm{max}}}$ & 1.731 & $\mathrm{g/(g\cdot h)}$ \\
$k_S$                  & $5.340\times10^{-7}$ & $\mathrm{g/L}$ \\
$k_{\mathrm{SV}}$      & $1.053\times10^{-6}$ & $\mathrm{VU/g}$ \\
$n_1$                  & $1.000\times10^{-2}$ & -- \\
$m_S$                  & $1.232\times10^{-6}$ & $\mathrm{g/(g\cdot h)}$ \\
$k_{\mathrm{XV}}$      & $2.605\times10^{-4}$ & $\mathrm{VU/g}$ \\
$n_2$                  & $1.028\times10^{-1}$ & -- \\
$Y_{\mathrm{XS}}$      & $1.083\times10^{-1}$ & $\mathrm{g/g}$ \\
$Y_{\mathrm{LX}}$      & 2.204 & $\mathrm{g/g}$ \\
$m_L$                  & 1.910 & $\mathrm{g/(g\cdot h)}$ \\
$k_{\mathrm{LS}}$      & $1.0\times10^{-10}$ & $\mathrm{g/L}$ \\
$k_{\mathrm{LV}}$      & 10.02 & $\mathrm{VU/g}$ \\
$n_3$                  & 10.0 & -- \\
$q_{E_0}$              & $1.000\times10^{-6}$ & $\mathrm{VU/(g\cdot h)}$ \\
$q_{E_{\mathrm{max}}}$ & 10.0 & $\mathrm{VU/(g\cdot h)}$ \\
$k_l$                  & $3.729\times10^{2}$ & \textmu$\mathrm{mol\, m^{-2}\, s^{-1}}$ \\
$n_4$                  & 4.718 & -- \\
$k_d$                  & 0.988 & $\mathrm{1/h}$ \\
\hline
$S_0$                  & 4.0   & $\mathrm{g/L}$ \\
$X_0$                  & 0.075 & $\mathrm{g/L}$ \\
$L_0$                  & 0.0 & $\mathrm{g/L}$ \\
$E_0$                  & 0.0 & $\mathrm{VU/g}$ \\
\hline
\end{tabular}
\end{table}

{In this second case study, we shifted the focus to tracking a defined intracellular ATPase trajectory, exemplifying a \textit{golden batch}. As such, the return function to maximize (cf. RL problem in \eqref{eq:rl_problem}) was:}
\begin{equation} 
{\mathbb{E}{\left[ J(\cdot) \right]} = \mathbb{E}{\left[-  \sum_{t=1}^{N_x} (E_t - E_{{r}_t})^2 \right]},} \label{eq:J_c2} 
\end{equation}
{where $E_{{r}_t}$ is the target reference at time $t$. The selected trajectory shape was derived from open-loop optimization experiments, as outlined in \citep{espinel-rios_experimentally_2024}. As will be shown, this golden-batch ATPase trajectory enables efficient management of metabolic trade-offs. In particular, high growth at low ATPase expression and increased lactate synthesis at high ATPase expression toward maximizing product titer in a given batch time. In Eq. \eqref{eq:J_c2}, we consider the \textit{negative} of the summed squared tracking error since the RL framework maximizes the expectation by default. Thus, the negative sign leads the agent to effectively minimize the tracking error. For comparison, our benchmark static metabolic control strategy was chosen to be a scenario with high ATPase expression, imposed by applying a constant input signal $u = 873$, the input's upper limit in \citep{espinel-rios_experimentally_2024}. This represents the case in which ATPase is expressed at its maximum level without any \textit{braking} mechanism.}

{\textit{Remark on RL training}. The same RL configuration and training aspects discussed in the fatty acid case study apply here. However, the input range was defined as $[0, 873]$, following the experiments by \citep{espinel-rios_experimentally_2024}. To improve stable convergence, we set the learning rate to $\alpha = 0.001$, and increased the allowed training epochs to 500, while keeping the early stopping strategy in place. In addition, we considered eleven equidistant stepwise constant inputs, applied over a process time of 9.5 h.}

\subsubsection{{Deterministic dynamics}}
\begin{figure}[htb!]
    \begin{center}
        \subfigure[\,\textbf{Return function}]{\includegraphics[scale=0.38, trim={0 10 0 10}, clip]{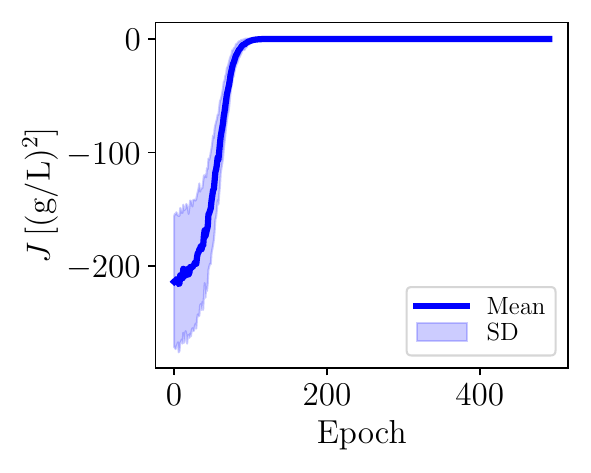}} 
        \subfigure[\,\textbf{Input}]{\includegraphics[scale=0.38, trim={0 10 0 10}, clip]{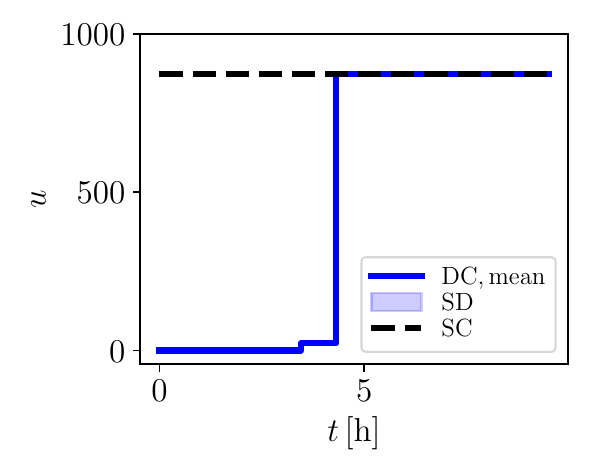}} 
        \subfigure[\,\textbf{Cell density}]{\includegraphics[scale=0.38, trim={0 10 0 10}, clip]{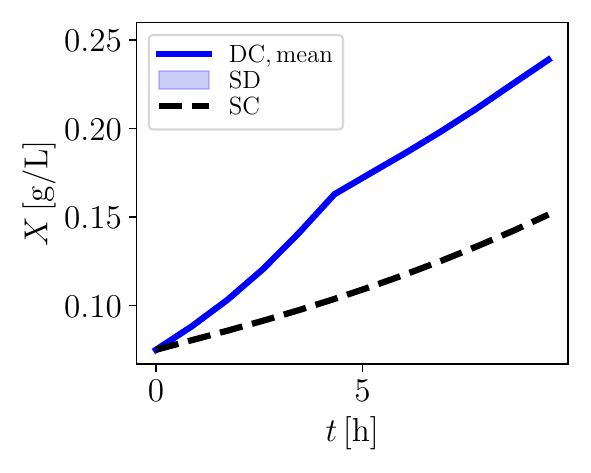}} 
        \subfigure[\,\textbf{Glucose}]{\includegraphics[scale=0.38, trim={0 10 0 10}, clip]{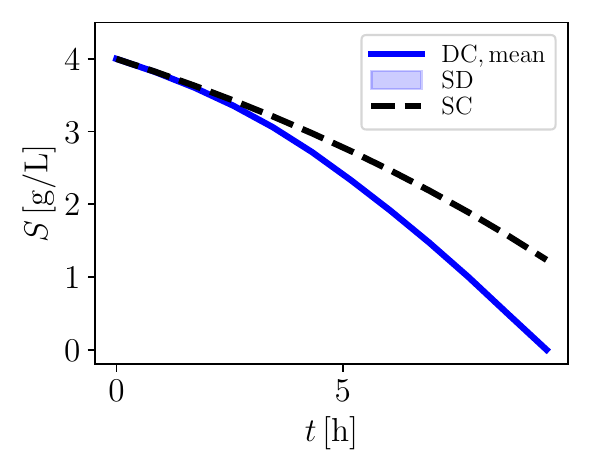}} 
        \subfigure[\,\textbf{ATPase}]{\includegraphics[scale=0.38, trim={0 10 0 10}, clip]{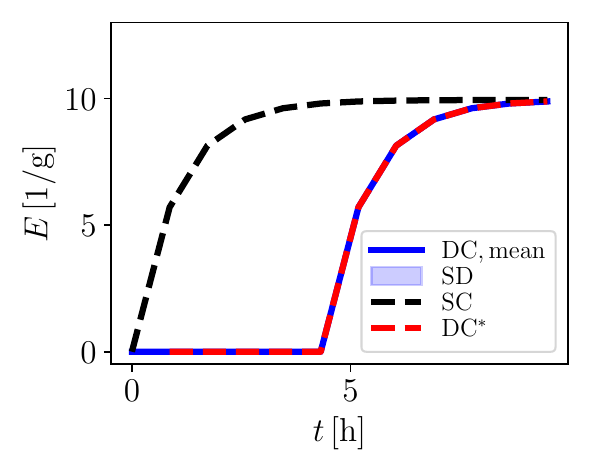}} 
        \subfigure[\,\textbf{Lactate}]{\includegraphics[scale=0.38, trim={0 10 0 10}, clip]{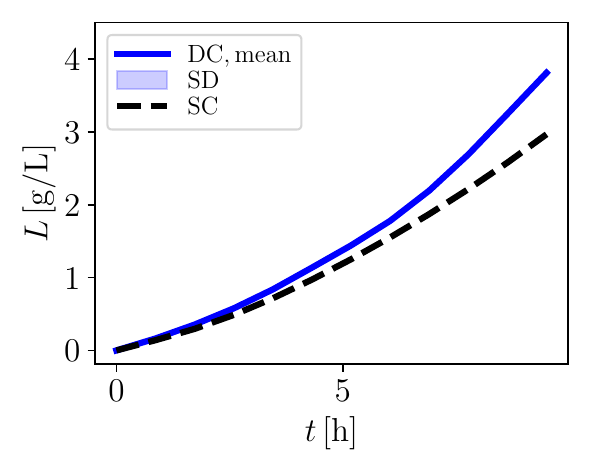}}         \caption{{Metabolic control results under ideal conditions (i.e., no system uncertainties) for the lactate biosynthesis case study with ATPase modulation. (a) Evolution of the return function over epochs, up to the epoch with the highest mean value (selected control policy). The corresponding (b) input trajectory and (c)-(f) dynamic state trajectories associated with the selected control policy are also shown. The RL-derived dynamic control scenario (DC) is benchmarked against the static control scenario (SC). The golden-batch ATPase trajectory is indicated with a red dashed-line (DC\textsuperscript{*}) in the ATPase plot. {Uncertainty bands correspond to 500 episodes or trajectories.} SD: standard deviation. {$J$: return, $u$: control input (inducer); $X$: biomass; $S$: glucose; $E$: manipulatable enzyme (ATPase); $L$: lactate.}}}
        \label{fig:0_unc_c2}
    \end{center}
\end{figure}

{Fig. \ref{fig:0_unc_c2} shows the performance of the RL-driven dynamic metabolic control strategy under ideal deterministic conditions (i.e., 0 \% uncertainty). The exploration-exploitation behavior was as anticipated; the RL agent explores more widely over the initial phase (i.e., larger standard deviation in the return), followed by more deterministic performance at later epochs.}

{The RL-derived dynamic metabolic control strategy leads to the successful reference tracking of the target ATPase dynamic trajectory. The predefined dynamic trajectory enables management of temporal trade-off between growth and enhanced lactate synthesis, resulting in a 28 \% higher final lactate titer and full substrate depletion, in contrast to the static control (fully-induced) approach. This is achieved through a switch-like input change introduced midway through the process. The outlined results demonstrate that, although the static metabolic control operation maximally increases the lactate synthesis rate, this is not necessarily optimal in terms of final product titer in the considered process timeframe due to the significantly impaired biomass growth.}

\begin{table}[h!]
\begin{center}
\caption{{Final lactate titers under static and dynamic control policies across different uncertainty levels in the lactate biosynthesis case study with ATPase modulation. Prediction uncertainty corresponds to 500 episodes or trajectories.}} \label{tab:summary_results_c2}
\begin{tabular}{|c|c|c|c|}
    \hline
    Unc. [\%] &  SC [$\mathrm{g/L}$] & DC [$\mathrm{g/L}$] & Imp. [\%]\\
    \hline
    0  &  2.97 $\pm$ 0.00 & 3.81 $\pm$ 0.00 & 28 \% \\
    \hline
    5  &  2.96 $\pm$ 0.23 & 3.66 $\pm$ 0.21 & 24 \% \\
    \hline
    10  &  2.99 $\pm$ 0.47 & 3.45 $\pm$ 0.39 & 15 \% \\
    \hline
    12.5 &  2.95 $\pm$ 0.57 & 3.36 $\pm$ 0.47 & 14 \% \\
    \hline
    15 &  2.97 $\pm$ 0.63 & 3.35 $\pm$ 0.54 & 13 \% \\
    \hline
\end{tabular}\\
\smallskip\noindent
{SC: static control. DC: dynamic control. Imp.: improvement. Unc.: uncertainty level.}
\end{center}
\end{table}

\subsubsection{{Policy robustness via domain randomization}}
\begin{figure*}[htb!]
    \begin{center}
        \subfigure[\,\textbf{5 \% uncertainty}]{\includegraphics[scale=0.38, trim={0 10 0 10}, clip]{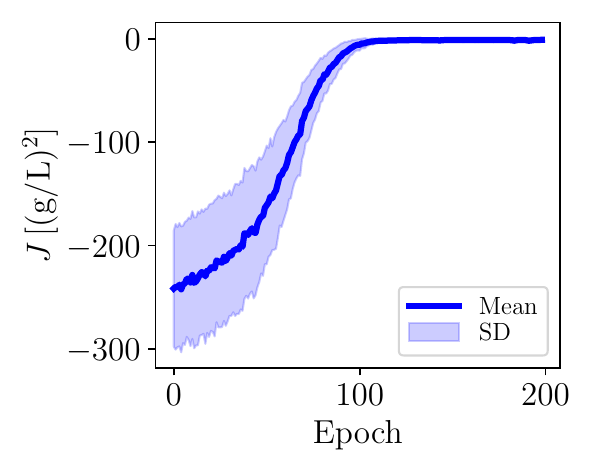}} 
        \subfigure[\,\textbf{10 \% uncertainty}]{\includegraphics[scale=0.38, trim={0 10 0 10}, clip]{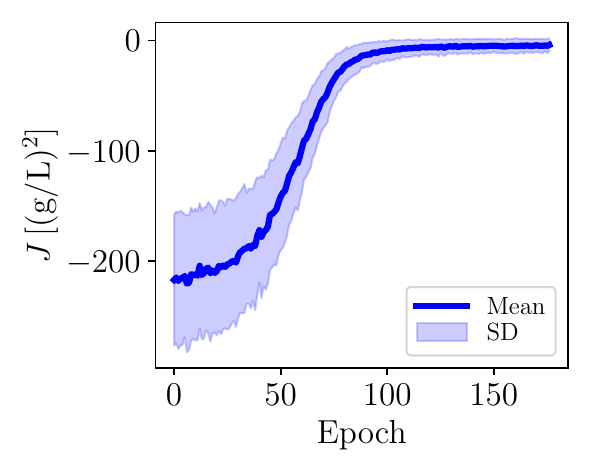}} 
        \subfigure[\,\textbf{12.5 \% uncertainty}]{\includegraphics[scale=0.38, trim={0 10 0 10}, clip]{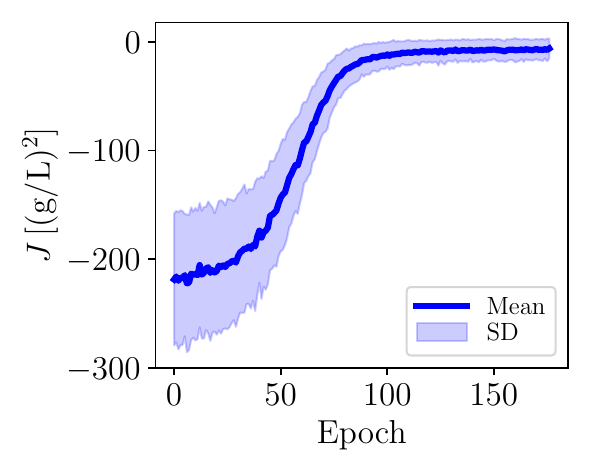}} 
        \subfigure[\,\textbf{15 \% uncertainty}]{\includegraphics[scale=0.38, trim={0 10 0 10}, clip]{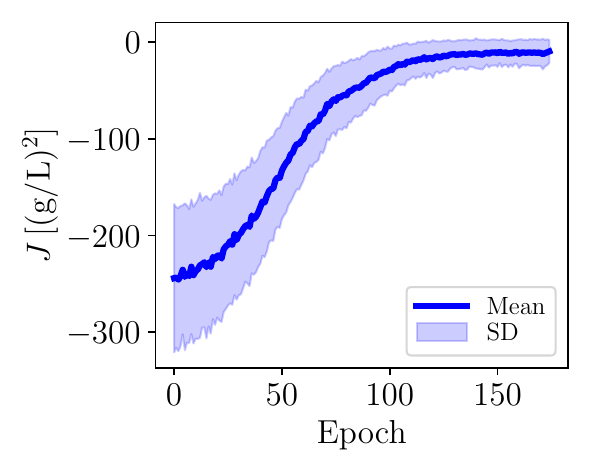}}

        \subfigure{\includegraphics[scale=0.38, trim={0 10 0 10}, clip]{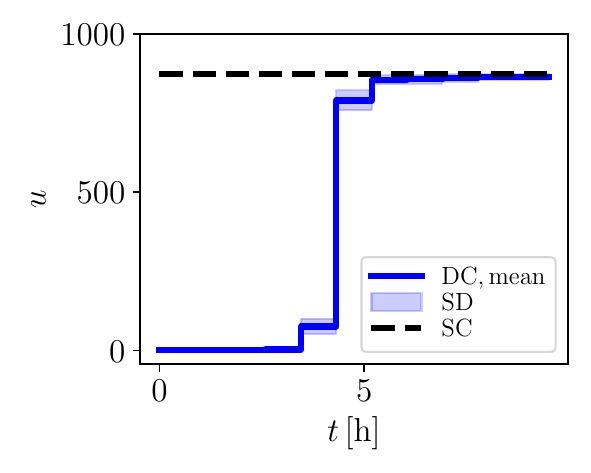}} 
        \subfigure{\includegraphics[scale=0.38, trim={0 10 0 10}, clip]{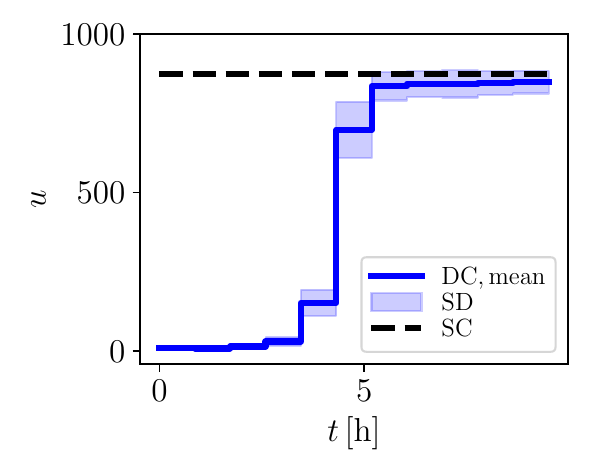}} 
        \subfigure{\includegraphics[scale=0.38, trim={0 10 0 10}, clip]{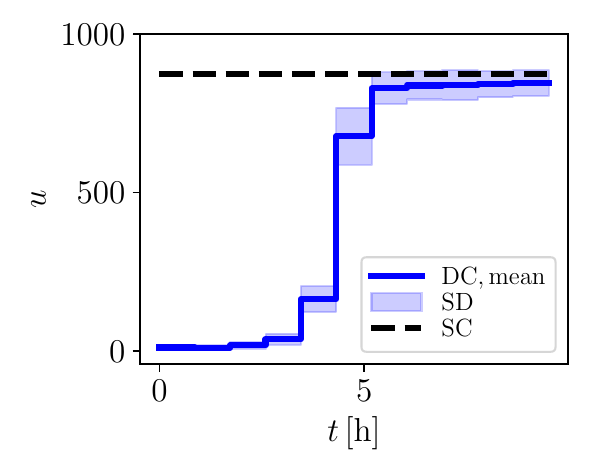}} 
        \subfigure{\includegraphics[scale=0.38, trim={0 10 0 10}, clip]{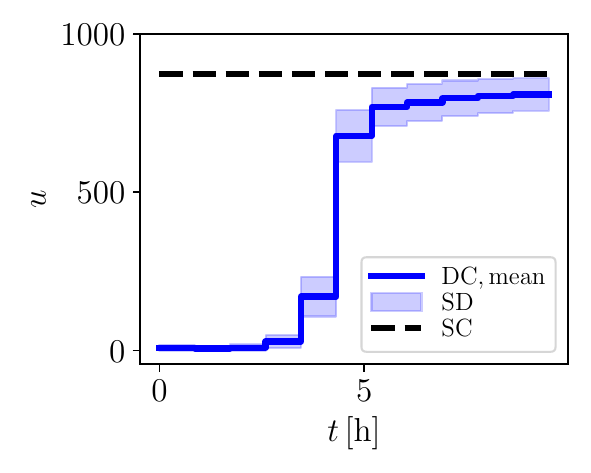}}

        \subfigure{\includegraphics[scale=0.38, trim={0 10 0 10}, clip]{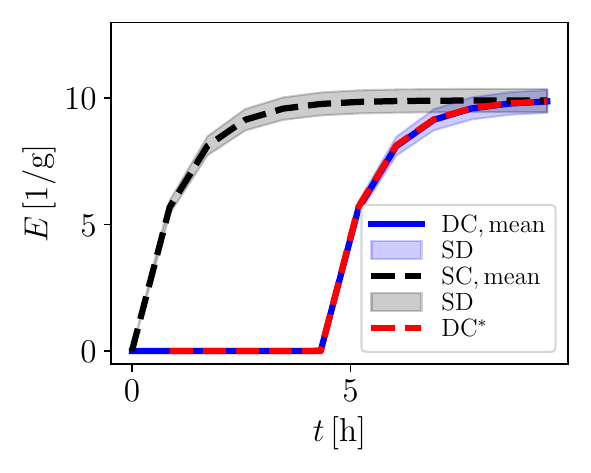}} 
        \subfigure{\includegraphics[scale=0.38, trim={0 10 0 10}, clip]{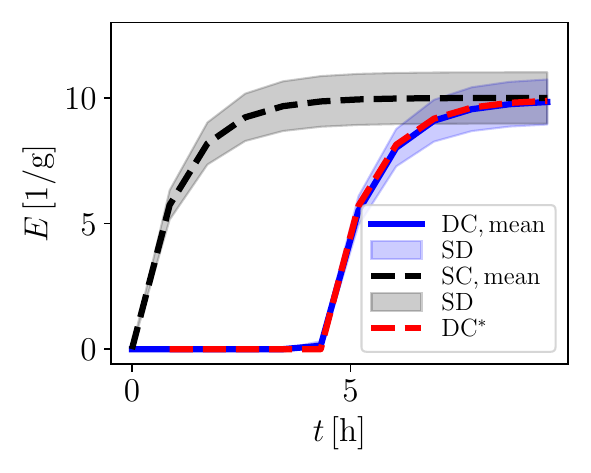}} 
        \subfigure{\includegraphics[scale=0.38, trim={0 10 0 10}, clip]{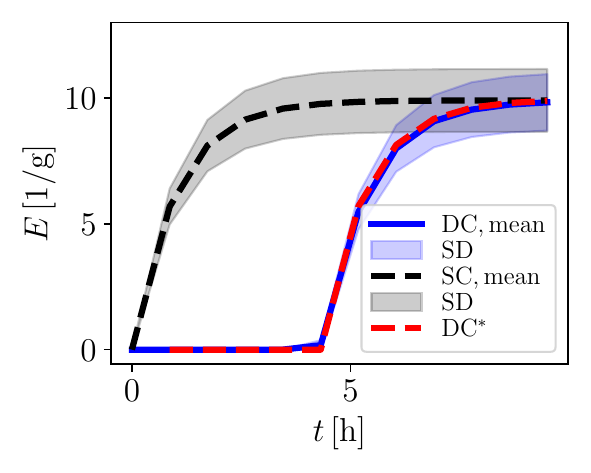}} 
        \subfigure{\includegraphics[scale=0.38, trim={0 10 0 10}, clip]{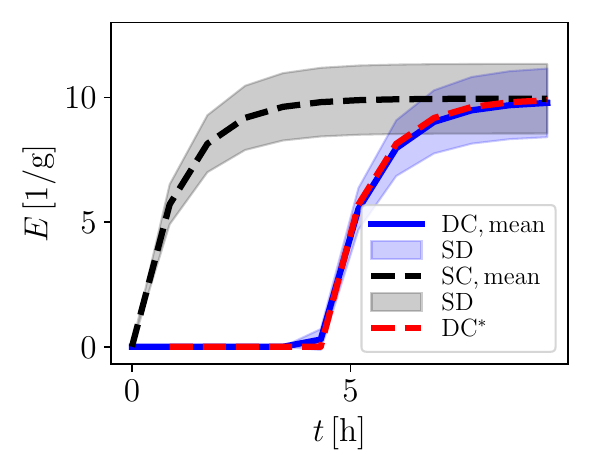}}

        \subfigure{\includegraphics[scale=0.38, trim={0 10 0 10}, clip]{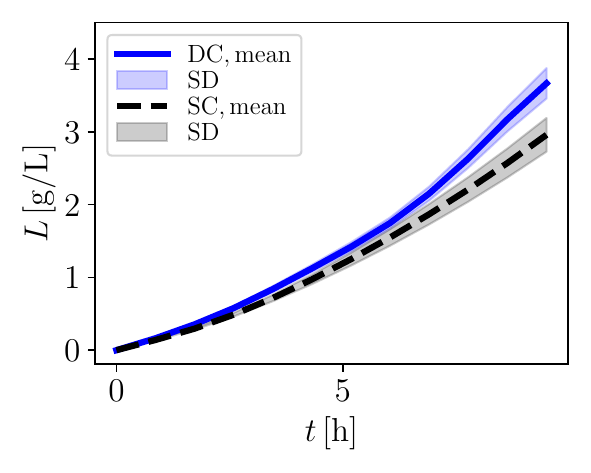}} 
        \subfigure{\includegraphics[scale=0.38, trim={0 10 0 10}, clip]{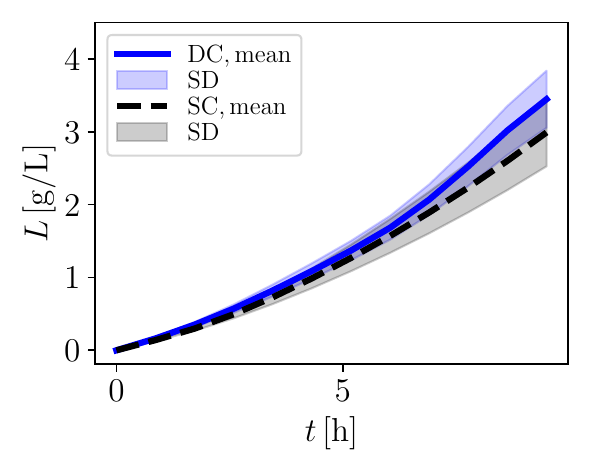}} 
        \subfigure{\includegraphics[scale=0.38, trim={0 10 0 10}, clip]{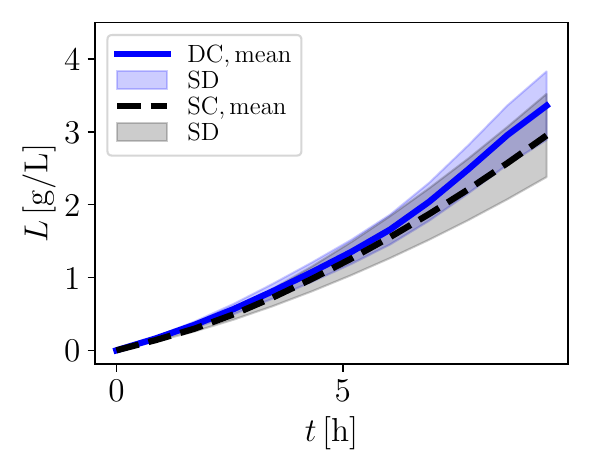}} 
        \subfigure{\includegraphics[scale=0.38, trim={0 10 0 10}, clip]{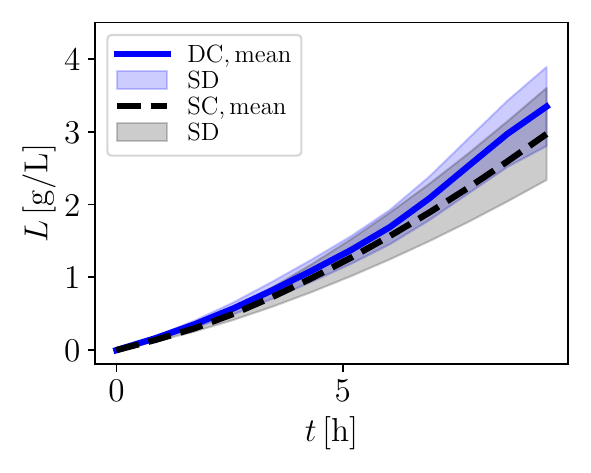}}
        
        \caption{{Control policies robust against system uncertainty for the lactate biosynthesis case study with ATPase modulation, considering (a) 5 \%, (b) 10 \%, (c) 12.5 \%, and (d) 15 \% uncertainty in the initial conditions and key kinetic parameters affecting the expression of ATPase. The RL-derived dynamic control scenario (DC) is benchmarked against the static control scenario (SC). The return function is presented up to the epoch with the highest mean return value, matching the chosen policy. Selected dynamic state trajectories correspond to the latter policy. The golden-batch ATPase trajectory is indicated with a red dashed-line (DC\textsuperscript{*}) in the ATPase plot. {Uncertainty bands correspond to 500 episodes or trajectories}. SD: standard deviation. $J$: return, $u$: control input (inducer); $X$: biomass; $S$: glucose; $E$: manipulatable enzyme (ATPase); $L$: lactate.}}
        \label{fig:fig_unc_c2}
    \end{center}    
\end{figure*}

{The performance of the RL-derived dynamic metabolic control policies {for the lactate biosynthetic system} under varying levels of uncertainty is shown in Fig. \ref{fig:fig_unc_c2}. A similar observation can be made with respect to the previously outlined fatty acid biosynthetic pathway. That is, higher uncertainty levels lead to greater standard deviations in both the return function evolution over epochs and the dynamic states for the best-performing epoch, reflecting the increased stochasticity in the bioprocess dynamics. Interestingly, the optimal input trajectory under stochastic conditions changes from a very steep switch-like pattern, in the deterministic setting, to a more gradual or smoother switching profile under uncertainty, thereby providing robustness.}

{Across all uncertainty scenarios, the dynamic metabolic control strategy consistently achieves mean lactate titer improvements of 13–28 \% relative to the static control approach \textit{under the same uncertainty conditions} (cf. Table \ref{tab:summary_results_c2}). Although the target ATPase trajectory was well-tracked in all dynamic metabolic control cases, a consistent decline in mean lactate titer is observed with increasing uncertainty, whereas the static control results remain mean-wise consistent throughout. The apparent robustness of the static control strategy can be attributed to the fact that induction remains at its maximum level at all times, regardless of uncertainty. In contrast, the dynamic control scenario involves more nuanced transient regulation, ranging from zero to (near-)maximum induction, making performance more sensitive to uncertainty in enzyme expression kinetics. Additionally, the reference trajectory was fixed and not adapted to account for system uncertainty. However, this fixed trajectory was intentional as the aim of this case study was to evaluate the RL agent’s ability to robustly track predefined golden-batch trajectories of manipulatable enzymes. This example represents cases where consistently reproducing intracellular behavior is of utmost importance to comply with pre-approved production protocols (e.g., involving regulatory entities) or to ensure that the cell manipulated within a \textit{safe} metabolic region, avoiding unstable states. Overall, the outlined lactate biosynthetic case study  further demonstrates the ability of our RL framework to work effectively in dynamic metabolic control contexts even under high system uncertainty.}

\section{Conclusions}
\label{sec:conclusions}
In this {study}, we proposed an RL-driven {framework to derive} dynamic metabolic control policies in bioprocesses. Our method {leverages} dynamic models as surrogate environments {and enhances robustness in the control policies through domain randomization. Domain randomization allows system stochasticity to be incorporated during RL training in a straightforward manner, making the learned policies uncertainty-aware}.

{As such, our framework provides a viable alternative to complex stochastic model-based control methods, such as stochastic MPC, which can be computationally demanding and challenging to implement, particularly for non-experts in control theory. Unlike model-based methods, which require differentiation with respect to decision variables, our outlined RL strategy only requires integrating the model forward in time; a much simpler task. When dynamic models exhibit highly nonlinear, stiff dynamics or piecewise kinetic functions with switch-like behavior or discontinuities, the convenience of our framework becomes even more evident.}

{The efficiency and robustness of our proposed RL framework was demonstrated using two biotechnologically relevant \textit{E. coli} bioprocesses as case studies. First, we maximized the product titer through optimal control of ACC expression in a fatty acid bioprocess. Then, we dealt with a trajectory tracking problem of a golden-batch ATPase trajectory in a lactate bioprocess. In both cases, we were able to efficiently derive robust dynamic metabolic control policies that successfully met the intended control objectives.}

{In addition, our framework enables, in principle,} the \textit{in silico} evaluation of different genetic circuit topologies in terms of control efficiency and robustness. This is particularly valuable in the early stages of research and development, where identifying the most promising {biocontrol} topologies and {manipulation strategies prior to experimental implementation is essential in order to save time and resources.}

{It is also worth noting that, in the current study, neural networks served as policy function approximators, trained to maximize a specified return. Therefore, we only sought efficient convergence with respect to the demanded return functions. Systematic policy generalization analysis, i.e., addressing policy overfitting or overspecialization, as well as experimental implementation in bioreactor setups, constitute ongoing work.}

\section*{{Author contributions}}
\label{sec:contributions}
{Sebastián Espinel-Ríos: Conceptualization; methodology; software; formal analysis; investigation; writing—review and editing; visualization. River Walser: Investigation (fatty acid case study, supporting); visualization (fatty case study, supporting). Dongda Zhang: Methodology; writing—review and editing.}

\section*{Acknowledgment}
\label{sec:acknowledgment}
\noindent SER is part of the Advanced Engineering Biology Future Science Platform (AEB FSP). {The authors thank Antonio del Rio Chanona for his valuable insights on reinforcement-learning concepts.}

\bibliography{bibliography}

\end{document}